\documentclass[a4paper,fleqn]{cas-sc}

\usepackage[numbers]{natbib}
\usepackage[section]{placeins}
\usepackage{xurl}
\makeatletter
\def\fps@figure{htbp}
\makeatother

\def\tsc#1{\csdef{#1}{\textsc{\lowercase{#1}}\xspace}}
\tsc{WGM}
\tsc{QE}

\begin{document}
\def\floatpagepagefraction{1}
\def\textpagefraction{.001}
\raggedbottom

\shorttitle{A Fully Automated DM-BIM-BEM Pipeline for Early Design}

\shortauthors{J. Xiao et al.} 

\title[mode = title]{A Fully Automated DM-BIM-BEM Pipeline Enabling Graph-Based Intelligence, Interoperability, and Performance-Driven Early Design}



%

\author[1,2]{Jun Xiao}

\author[1]{Qiong Wang}

\author[3]{Yihui Li}

\author[1]{Zhexuan Yu}

\author[4]{Hao Zhou}

\author[1,2]{Borong Lin}

\ead{linbr@tsinghua.edu.cn}

 \ead[url]{https://github.com/UmikoXiao}


\affiliation[1]{organization={Department of Building Science, School of Architecture},
	addressline={Tsinghua University}, 
	city={Beijing},
	postcode={100084}, 
	state={Beijing},
	country={China}}

\affiliation[2]{organization={Key Laboratory of Eco Planning \& Green Building, Ministry of Education},
	addressline={Tsinghua University}, 
	city={Beijing},
	postcode={100084}, 
	state={Beijing},
	country={China}}

\affiliation[3]{organization={Center for Environment Design Research},
addressline={University of California, Berkeley}, 
city={Berkeley},
postcode={94704}, 
state={California},
country={United States}}

\affiliation[4]{organization={Institute for Urban Governance and Sustainable Development},
addressline={Tsinghua University}, 
city={Beijing},
postcode={100084}, 
state={Beijing},
country={China}}

\cormark[1]












\cortext[cor1]{Corresponding author}



\begin{abstract}
Graph-based artificial intelligence methods are increasingly advocated for early-stage energy-efficient building design, yet fail to adopt flexible boundary-representation (B-rep) models due to the lack of explicit spatial structure and topology. This paper presents a fully automated framework transforming B-rep geometry into semantically enriched ontology-based knowledge graphs and executable building energy models. The workflow integrates automated geometry cleansing, multiple space-generation strategies, graph-based extraction of spatial and element topology, and reversible translation to simulation files. The framework is validated using three datasets comprising parametric models, sketch-based conceptual designs, and complex real-world buildings, demonstrating high robustness, accurate reconstruction of spatial topology, efficient space generation, and close agreement between simulated energy results and reference models. Implemented as an open-source plugin with direct integration into SketchUp environment or standalone python package, the framework enables automated transformation from design geometry to analysis-ready energy models, supporting scalable automation and data-driven workflows in early-stage building performance analysis.
\end{abstract}


\begin{highlights}
\item Fully automated transformation from B-rep design models to energy models
\item Validated on large-scale datasets including complex real-world buildings
\item High fidelity topology extraction for early-stage building analysis
\item Graph-based BIM enabling AI-driven design and performance simulation
\item Higher robustness than commercial tools for complex and irregular geometries
\end{highlights}


\maketitle

\begin{table}[htbp]
	\centering
	\caption{Nomenclature}
	\label{Nomenclature}
	{ 
		\begin{tabular}{l l l l}
			\toprule
			Symbol & \multicolumn{1}{l}{Description} & Abbreviation & Full Name\\  
			\midrule
			$(S_{}^{ab})$ & Similarity of node a and b & DM & Design Model\\
			\midrule
			$s_N(g_a,g_b)$ & Gross node similarity of graph a and b &BIM&Building Information Model\\
			\midrule
			$s_{NA}(g_a, g_b)$ & Space similarity of graph a and b &BEM&Building Energy Model\\
			\midrule
			$GED$ & Graph Edit Distance &GNN&Graph Neural Networks\\
			\midrule
			$nGED$ & negative Graph Edit Distance &OWL& Web Ontology Language\\
			\midrule
			$E(g_a)$ & Processing efficiency of graph a &KG& Knowledge Graph\\
			\midrule
			$sC$ & Robustness (\% of Successful cases) &LPG& Label Property Graph\\
			\midrule
			$sA$ & Gross Floor Area accuracy of a case &1LSB&1st Level Space Boundary\\
			\midrule
			&&2LSB&2nd Level Space Boundary\\
			\midrule
			&&ASG&Auto Space Generation\\
			\midrule
			&&BTG&Building Topology Graph\\
			\midrule
			&&CCR&Close Contour Recognition\\
			\midrule
			&&VFG&View Factor Graph\\
			\bottomrule
		\end{tabular}
	}
	\\[1ex] 
\end{table}

\section{Introduction}
\label{sec:introduction}
Most high-sensitivity building performance parameters are defined at the early design stage, it is estimated that more than 40\% of the potential energy savings can be preserved \cite{Warren2002}. Architects and designers—key stakeholders are increasingly motivated to embed efficient performance analysis and optimization directly into early design workflows requiring high efficiency and flexibility to fit the frequent design iteration \cite{Purup2020}. Consequently, this has spawned the potential and demand for AI intervention in preliminary design.

One of the main barriers for AI intervention is the fully automated pipeline for the model representation transformation. For the perspective of encoding on spatial information and building properties, especially thermal properties highly related to the building performance, the Building Information Models (BIM) represented by Knowledge Graph (KG) is one of the best information carrier for AI interpretation. Integrating KG structures into automated workflows enables advanced applications such as language representation learning \cite{Ji2022}, question answering, and recommender systems \cite{Li2026BuildingGPT}. Label Property Graph (LPG)-based representations further support node and edge prediction using graph neural networks (GNNs), including neighborhood-scale energy prediction for building networks \cite{Wu2025}. Specifically, the automatic pipeline for AI intervention on design and building performance is hindered by two reversible transformation: (1) converting design models to structured Building Information Models (BIM), and (2) converting BIM to detailed BPS analytical models (BPAM). They have different representation and information aiming on their application scenarios: 

\begin{itemize}
	
\item \textbf{Design Models (DM)} capture only the geometric configuration of buildings, typically using boundary‐representation (B-rep) schemas (faces, edges, normals, materials) as in SketchUp or Autodesk Maya, or Constructive Solid Geometry (CSG) using volumetric primitives. Early-stage DMs intentionally remain minimal—lacking semantic structure or thermal attributes—to maximize flexibility during conceptual design.

\item \textbf{Building Information Models (BIM)} formalize spatial hierarchies, element attributes, and semantic relationships. They are frequently encoded in the Web Ontology Language (OWL) through domain ontologies such as BOT \cite{Rasmussen2025} or BRICK \cite{Brick2025}, with the IFC schema (*.ifc) and its ifcOWL representation \cite{buildingSMART2025ifcOWL} serving as the dominant standard. These OWL-based BIMs can be represented as Knowledge Graphs (KGs) consisting of RDF triplets, or equivalently as LPG of nodes and edges.

\item \textbf{BPS Analytical Models (BPAM)} extend BIM by embedding simulation-specific parameters into spaces and building elements. Required information varies by engine: space-face-space topology for multi-zone energy or airflow analysis, system topology for system-dynamics tools (e.g., TRNSYS, VENSIM), and categorized meshes or volumetric partitions for daylighting or CFD. This work focuses on the Building Energy Model (BEM), the most information-intensive form of BPAM.

\end{itemize}

In the DM-to-BIM transformation, a B-rep model must be structurally enriched with spatial hierarchies and augmented with topology describing relationships among building elements. This requires computing two key topological structures: the space-face topology represented as the 1st Level Space Boundary (1LSB) \cite{buildingSMART2023IfcRelSpaceBoundary1stLevel}, and the face-face topology represented as the 2nd Level Space Boundary (2LSB) \cite{buildingSMART2023IfcRelSpaceBoundary2ndLevel}. Existing research (Table~\ref{tab:DM2BIM}) focuses predominantly on 2LSB computation within OWL-based BIM environments (e.g., *.ifc), but these methods presuppose an available 1LSB and thus cannot support early-stage B-rep models that lack explicit space definitions. In contrast, 1LSB generation—Auto Space Generation (ASG)—remains underdeveloped: current methods assume idealized geometries, simplified spatial layouts, and limited model diversity. Consequently, a unified transformation framework is needed—one that robustly computes 1LSB from heterogeneous, unstructured B-rep inputs, while remaining compatible with existing 2LSB algorithms.

\begin{table}[htbp]
	\centering
	\caption{Input Requirements and Corresponding Outputs of Relevant Studies}
	\label{tab:DM2BIM}
	{ 
		\begin{tabular}{l l l l l}
			\toprule
			Ref. & \multicolumn{1}{l}{Input Requirement} & \multicolumn{3}{l}{Output} \\  
			\cmidrule(lr){2-5}  
			& Element \& Geometries & Efficiency & 1LSB & 2LSB \\  
			\midrule
			\cite{Rose2015} & \shortstack[l]{No gaps and clashes \\ Classified Elements}  & 3.39s/space & N/A & Face-Space topology (IFC2x3) \\
			\midrule
			\cite{Ladenhauf2016} & \shortstack[l]{No gaps and clashes \\ Classified Elements} & N/A & N/A & \shortstack[l]{Face-Face topology \\ Face-Space topology} \\
			\midrule
			\cite{Ying2021Generating} & \shortstack[l]{No gaps and clashes \\ Classified Elements}  & 0.10s/Space & N/A & \shortstack[l]{Face-Face topology \\ Face-Space topology (IFC4)} \\
			\midrule
			\cite{Ying2021rule} & \shortstack[l]{No gaps and clashes \\ Classified Elements}  & 7.19s/Space & N/A & \shortstack[l]{Face-Face topology \\ Face-Space topology (IFC4)} \\
			\midrule
			\cite{Chen2018graph} & \shortstack[l]{no gaps and clashes \\ fully aligned floors and walls \\ Closed loop of walls} & 0.17s/Space & 3d-1LSB & Face-Space topology \\
			\midrule
			\cite{LJones2013} & \shortstack[l]{Geometry with minimal gaps \\ only space boundary element} & 3.10s/Space & 3d-1LSB & Face-Space topology \\
			\midrule
			\cite{Xiao2023} & \shortstack[l]{No gaps and clashes \\ Closed loop of walls} & 0.06s/Space & 3d-1LSB & Face-Space topology \\
			\bottomrule
		\end{tabular}
	}
\end{table}

In the BIM-to-BEM transformation, OWL-encoded BIMs must be converted into analytical models by embedding thermal and system parameters required for BPS. Prior research has examined conversions from BIM to EnergyPlus (*.idf) and to system-modelling environments such as Modelica, revealing two central challenges: (1) validating OWL-based BIM representations \cite{Li2024BIM}, and (2) transforming the validated knowledge graph into simulation-ready analytical models (Table~\ref{tab:BIM2BEM}). Some studies address these jointly by applying template-based population of missing thermal settings \cite{Wu2024knowledge}, thereby increasing compatibility with diverse OWL inputs. However, template-driven workflows remain inflexible, requiring users to select or customize OWL models to define thermal attributes. Additionally, heterogeneity in file formats and data requirements leads to inconsistent transformation pipelines that often depend on format checks or intermediary representations, limiting interoperability across BIM ontologies and simulation engines \cite{Li2024BIM}.

\begin{table}[htbp]
	\centering
	\caption{Input Requirements and Simulation Targets of Relevant Studies}
	\label{tab:BIM2BEM}
	\resizebox{\columnwidth}{!}{
		\begin{tabular}{l l l l l}
			\toprule
			Ref. & Input Requirement & Ontology & Sim. format & Sim. target \\
			\midrule
			\cite{Wu2024knowledge} & \shortstack[l]{1LSB/2LSB \\ Geometry \\ Semantic information} & PGD, BRICK, BOT & EnergyPlus & Sketch stage BPS \\
			\midrule
			\cite{Wang2024} & \shortstack[l]{2LSB \\ Geometry \\ HVAC system topology \\ Material properties} & IFC4, Brick Schema & EnergyPlus & \shortstack[l]{Indoor temperatur\\ energy consumption} \\
			\midrule
			\cite{Iliadis2025} & \shortstack[l]{2LSB \\ HVAC system topology \\ Material properties} & \shortstack[l]{IFC, IDS, \\Labelled Property Graph (LPG)} & \shortstack[l]{Modelica BuildSys,\\ AixLib libraries} &\shortstack[l]{Dynamic energy assessment, \\\shortstack[l]{baseline and renovation \\ scenario evaluation}} \\
			\midrule
			\cite{Sayegh2024} & \shortstack[l]{2LSB \\ Geometry \\ Material properties \\ Occupancy scenarios} & IFC4, gbXML & \shortstack[l]{Modelica BuildSysPro\\ library (Dymola)} & \shortstack[l]{Building energy model \\ digital twins, \\energy assessment} \\
			\midrule
			\cite{Jansen2021BIM2SIM} & \shortstack[l]{2LSB \\ Geometry \\ HVAC system topology \\ Material properties} & IFC, MVD & \shortstack[l]{Modelica \\(AixLib, HKESim), \\EnergyPlus, CFD} & \shortstack[l]{BEPS\\ HVAC simulation, CFD, \\Life Cycle Assessment (LCA)} \\
			\midrule
			\cite{Andriamamonjy2018} & IFC4 with IDEAS library & IDM, MVD & Modelica & \shortstack[l]{building with \\ Modelica Library} \\
			\midrule
			\cite{Pinheiro2018MVD} & \shortstack[l]{2LSB \\ Geometry \\ HVAC system topology \\ Material properties \\ Occupancy scenarios} & IFC4, IDM, MVD & \shortstack[l]{EnergyPlus \\ Modelica \\(AixLib, BuildSys)} & Generic energy simulation \\
			\bottomrule
		\end{tabular}
	}
\end{table}

Given the persistent format-translation barriers, existing methods still lack a fully automated workflow from DM to BEM, especially when starting from unstructured B-rep design models. Current pipelines do not provide a unified or robust architecture capable of accommodating heterogeneous ASG and 2LSB methods, and data structures remain poorly aligned across the two transformation stages. To overcome these limitations, this article develops a comprehensive DM→BEM framework with three key contributions:

\begin{itemize}

\item \textbf{Fully automated and reversible end-to-end pipeline.}
The proposed framework establishes a complete automated workflow that unifies geometry cleansing, multi-strategy first-level space boundary (1LSB) generation, 2LSB topology extraction, and BEM generation within a single pipeline. The system performs end-to-end conversion—from DM to OWL-based BIM and finally to a BEM in EnergyPlus (*.idf)—while maintaining reversibility between stages. Each module is designed to be interchangeable, enabling plug-and-play integration of alternative algorithms and facilitating extensibility for future research.

\item \textbf{Broad format support with ontology-consistent alignment.}
To ensure data coherence across transformations, a unified data structure centered around an intermediate OWL representation is designed to resolve data-schema misalignments across transformation stages. The OWL model is aligned with major building ontologies—including IFC, BOT, and Brick—allowing consistent semantic representation and seamless translation between heterogeneous formats. This connection bridge DM and BEM to allow the framework accepts multiple common B-rep formats (*.obj, *.stl, etc.) and generates an intermediate OWL model aligned with major ontologies (ifc4.0, gbXML, BOT, WGS, etc.), enabling seamless downstream transformation through existing ontology-alignment modules.

\item \textbf{Systematic implementation, quantified validation, and benchmarking of ASG and BIM-to-BEM methods.}
This framework allow the embedding of traditional ASG or BIM-to-BEM methods following the data alignment restriction. In this research, three distinct ASG strategies are systematically reproduced, optimized, and evaluated within the framework. This unified implementation enables direct comparative benchmarking of their performance in terms of robustness, geometric accuracy, and computational efficiency when applied to diverse and imperfect B-rep design models. The resulting evaluation provides one of the first consistent performance assessments of ASG methods under realistic modelling conditions. For the sake of practical scenario, accuracy, robustness, and computational performance have been evaluated against comparable software using three datasets: 100 procedurally generated B-rep models, 570 manually created models from architecture students, and 15 complex real-building cases, with all metrics systematically assessed.

\end{itemize}

This framework enables broad practical applications. For DM→BEM, it supports direct energy simulation on arbitrary early-stage models. For DM→BIM, it extends KG-based design analytics—such as semantic interpretation, automated reasoning, and optimization—to flexible conceptual geometry. For BIM→BEM, it provides a unified mechanism for translating between mesh-based (e.g., RADIANCE, CFD) and topology-based (e.g., AFN, EnergyPlus) simulations, enabling integrated multi-method building performance analysis.

\section{Method}
\begin{figure}
	\centerline{\includegraphics[width=\columnwidth]{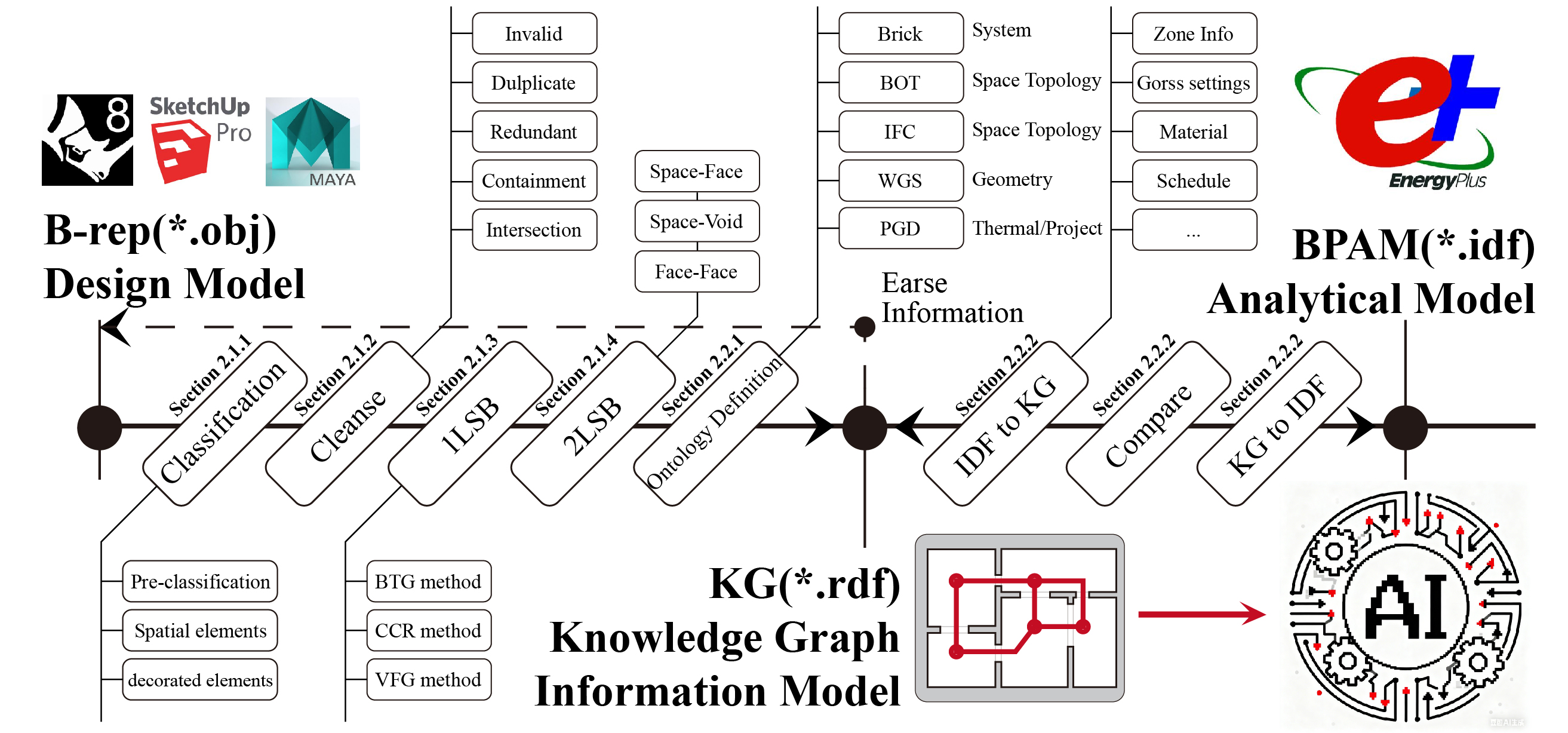}}
	\caption{Modules in the transformation framework. Each module is a stand-alone method and allows substitution following the API documentation on the project}
	\label{figFramework}
\end{figure}

The workflow of the framework is generally serialized in two transformation tasks divided into eight modules (Fig.\ref{figFramework}).

\subsection{Design Model to Building Information Model Transformation}
This transformation provides a module that converts a design-oriented, stream-based B-rep file (*.geo) into an Information Graph file (*.owl or *.gbxml). The framework is also compatible with widely used B-rep formats through a translation module for *.stl and *.obj files, and automatically performs geometric cleansing during the transformation process.

The framework has developed an iterative geometry-processing module to support complex Boolean and 3D calculations based on the Python package PYGEOS. This package has high efficiency and robustness, but only supports basic calculations on 2D geometries.

This framework has made a key input assumption that the framework expects building elements like walls to be represented as single surfaces, which is a common practice in early-stage conceptual modeling.

\subsubsection{Face Classification}
The Face classification is one of the main enrichment process from DM to BIM. It has three elaborated steps during the whole transformation to encode those faces into elements:
	
\begin{itemize}
	\item \textbf{Pre-classification:}
	A robust classification highly rely on accurate 1LSB However, most of the 1LSB calculation method require a raw identification on horizontal faces and wall-liked vertical faces. In this case, a pre-classification before 1LSB calculation would operate based on face attributes, transparency, and normal. Besides, the aperture (Door-like, window-like, skylight-like faces, etc.) would be attached to the parent opaque faces based on the 2d projection and faces' normal, then removed from the 1LSB calculation; those do not find an attachment would be regarded as curtain-like faces and maintained in the 1LSB process. In this process, the aperture elements would be encoded. This framework allow the connections with other pre-classification method, and tag the primary identification of the faces in the stream-based B-rep file (*.geo) following our instruction (where the pre-classification process will be skipped).
	\item \textbf{Spatial elements identification:}
	After the 1LSB is provided, the faces composing 1LSBs would be formatted into fusiform structures \cite{Chen2018graph} based on the connectivity. In this process the ceiling, wall and floor elements would be identified, and a cleanse process (see solve\_intersection in Section 2.1.2) would operate to ensure the tightness between ceiling/floor and wall. In this process, the ceiling, floor and wall elements would be encoded.
	\item \textbf{decorated elements attachment:}
	 After filtering out faces that do not contribute to the spatial structure, the process produces more detailed face tags (e.g., shading, extra content, internal mass, etc.) for use in later stages of the workflow. These faces would try to find an attachment to the connected aperture as a shading elements; otherwise, the faces would be regarded as internal mass, extra wall, and floor elements.
\end{itemize}

Currently the stairs or other detail elements could not be recognized and transformed. This leads to the low sA of some of the hand-made sketch model in datasets 2. In this case, we suggest users do not include too many non-spatial and non-structural decorative elements except for internal walls/floors and shadings.

\subsubsection{Cleanse}
\begin{figure}
	\centerline{\includegraphics[width=\columnwidth]{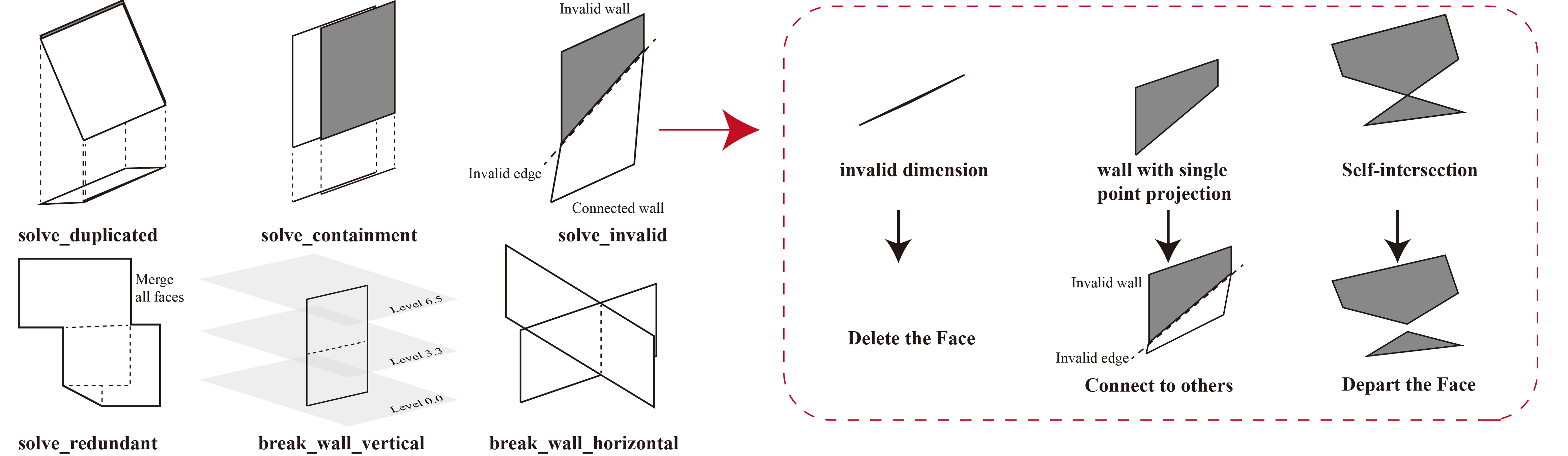}}
	\caption{visualization of 6 cleansing methods}
	\label{cleanseMethod}
\end{figure}

The cleansing process is a critical component for ensuring the robustness of the transformation. In this framework, the data-cleansing procedure comprises six primary sub-modules (Fig. \ref{cleanseMethod}):

solve\_duplicated (Optional): Identifies and removes faces with identical levels, 2D projections, and heights. Faces are considered duplicates when their 2D projections coincide and all coordinates fall within a prescribed threshold.

solve\_redundant (Optional): Detects and merges co-planar faces. This step is essential for mesh-based inputs (e.g., *.stl). An edge dictionary is constructed to identify shared edges, allowing faces with identical normals to be merged.

solve\_intersection (Optional): Identifies overlapping or enclosed co-planar faces and the intersection between vertical and horizontal faces. Containment is detected when faces on the same level have intersecting 2D projections; and the Containment between 2D-1LSB and the ceiling/floor-like faces is detected with the same algorithm after the 1LSB calculation. This design can avoid calculating heavy 3D intersection with the assurance of tightness.

solve\_invalid: Removes invalid faces, including those with degenerate projections, self-intersections, or incorrect geometric dimensions.

break\_wall\_vertical (Optional): Splits walls spanning multiple building levels to support multi-level thermal zone identification. Walls with vertical extents exceeding two levels are segmented.

break\_wall\_horizontal (Optional): Subdivides walls at their intersections with other walls. This is required for B-rep models lacking precomputed face intersections, ensuring proper segmentation of vertical faces.

The cleanse module is not only applied after the geometries input. To enhance the robustness of the whole transformation, especially the 1LSB calculation, these modules were highly combined with the classification process. Before the pre-classification, solve\_duplicated, solve\_redundant, break\_wall\_vertical methods (during vertical faces identification) would be applied to ensure the unique attachment of the aperture. After that, all cleanse would be applied to ensure a clean wall-like and face-like faces set for the 1LSB calculation. Before extruding the 2LSB topology, the solve\_intersection would be recomputed to ensure the tightness of the floor and ceiling elements. Besides, after each cleanse, solve\_invalid would be applied to erase the invalid elements or faces.

\subsubsection{First-Level Space Boundaries (1LSBs) Generation}
The 1LSBs define the envelope surfaces of a space in a fusiform structure \cite{Lin2021MOOSAS}, consisting of a floor (potentially composed of multiple planes), a roof (also with multiple planes), and a 2D boundary formed by a closed loop of walls. This process encounters the main technical difficulty in the Auto Space Generation (ASG) question, with physical or topological solutions.

The physical ASG solution approximated the close volume based on real locations of the faces. These methods is advanced in their tolerance to the physical disjoint like the gaps create by manual modeling. These methods can flexibly adjust the tolerance size in the boolean algorithm. For example, the View Factor Graph (VFG) method \cite{LJones2013} could adjust the density of rays or maximum distance in intersection testing to allow small creaks between walls; other pixel-based space identify method \cite{CVPRFloorASG} tag the pixels and calculate the 2d-1LSB, which can adjust the pixel size to meet the tolerance.

The topological ASG solution will first solve the connectivity between elements to build a undirected graph, and try to calculate the 1LSB directly on the graph. These methods are advanced in their efficiency exceeding the scale limitation on the physical size of the model; at the same time avoid boolean calculation on geometries which are very time-consuming. The main difficulties of these methods lay on the disjointed spatial volumes create by manual modeling. In this case the algorithm should first fusion near vertex and prune redundant edges to allow tolerance on the physical disjoint to improve the graph quality and 1LSB accuracy.

\begin{figure}
	\centerline{\includegraphics[width=\columnwidth]{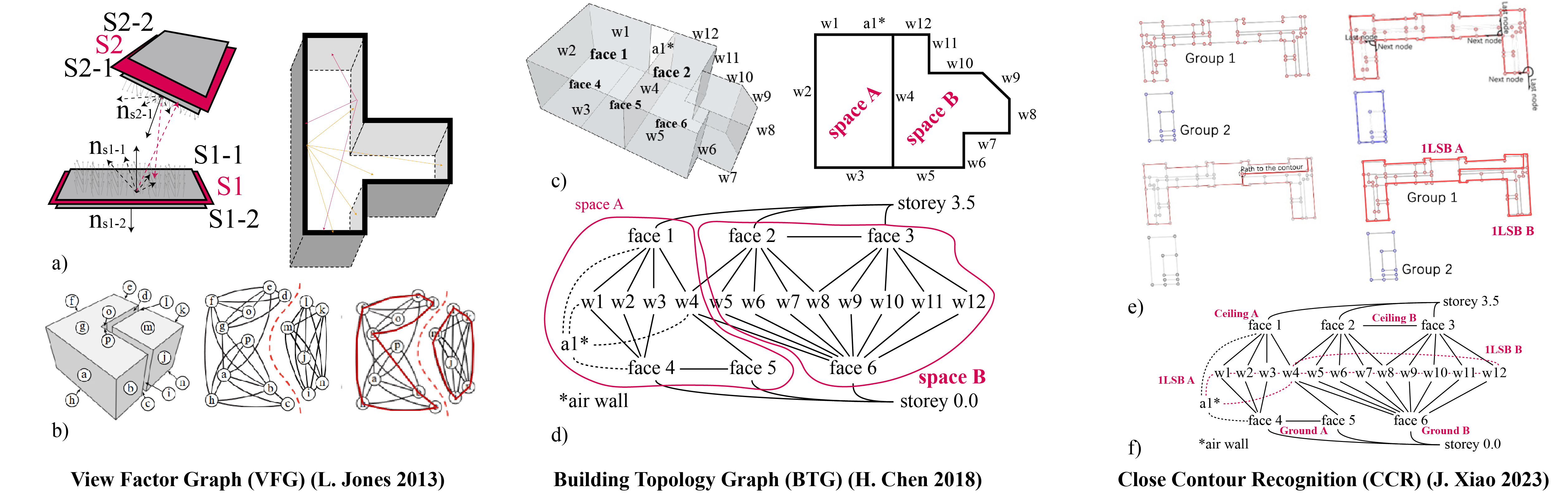}}
	\caption{ASG methods description. a) The view factors to link elements, each element has 2 surfaces; b) Split the connection network on elements and calculate the envelope; c) The topology and space relations to identify by BTG; d) The fusiform structure of two spaces; e) The recursively division logic of the CCR method; f) Boolean and location of the ceilings and grounds.}
	\label{ASG}
\end{figure}

Three representative ASG methods capable of computing 1LSBs have been reproduced, optimized, and integrated into this framework (Fig. \ref{ASG}), including the physical method VFG and topological method BTG and CCR. This framework does not include the pixel-based method since their program structure is totally different from ours, requiring huge efforts in I/O and data alignment.

The View Factor Graph (VFG) method \cite{LJones2013} computes visibility between surfaces to assemble closed fusiform spaces. On each side of a surface, the algorithm first randomly creates rays from points on the face to test the visibility of this one-side surface (Fig. \ref{ASG}-a). An undirected graph representation would be built based on the visibility between faces; then the Label Propagation Algorithm (LPA) is used to partition the connected groups in the graph into isolated node clusters. In Fig. \ref{ASG}-b, the 1LSB would be generated based on the connectivity within the cluster. It requires substantial ray-mesh Boolean computation but allows accuracy-efficiency trade-offs by adjusting ray density. Moreover, it remains robust when edge gaps between faces are smaller than the ray solid angle.

The Building Topology Graph (BTG) method \cite{Chen2018graph} constructs fusiform structures directly from face-face connectivity. This method uses any horizontal faces as floor faces, and directly builds a 2d-1LSB with a loop of walls (Fig. \ref{ASG}-c). Those "walls" would find exact vertical faces to attach, otherwise they would be regarded as air boundary. Then, space would be connected according to the shared air boundary as the final 1LSB. In Fig. \ref{ASG}-d, a similar method has been applied in Sefaira \cite{Sefaira2023} for SketchUp DM transformation. Although computationally efficient, BTG struggles with complex geometries such as vertically curved walls (walls composed of multiple vertical faces) or atria extending across multiple storeys.

The Close Contour Recognition (CCR) method \cite{Xiao2023} identifies 2D 1LSBs using a dimension-reduced model in which faces are assigned to storeys, and then generates 3D 1LSBs through Boolean operations between 2D boundaries and planes. All walls would be projected to the 2d plane according to their bottom edge. The close contour, that is, the 2d-1LSB, would be calculated on the connectivity graph of walls in each building level: firstly, partition the graph by LPA into isolated clusters; then, the outer boundary of each cluster would be extracted. Finally, those boundaries would be recursively divided until no extra walls are located in the boundaries. In Fig. \ref{ASG}-e, the 2d-1LSB would be formed into the final 1LSB by attaching floors and ceilings. This approach, similar to AutoCAD’s hatch-area identification, is computationally more intensive but offers greater robustness for storey-based architectural designs.

\subsubsection{Second-Level Space Boundaries (2LSBs) Generation}
This stage extracts the topological relationships between identified thermal zones and their bounding insulation elements (2LSB), thereby forming a complete multi-zone space network. The 2LSB definitions follow the IFC 4.x ontology \cite{buildingSMART2023IfcRelSpaceBoundary2ndLevel}. Any 2LSB computation method can be integrated into the framework, provided it adheres to the class definitions of elements (WALL, FACE, GLAZING, SKYLIGHT) and spaces (SPACE, VOID).

Numerous mature 2LSB generation approaches exist (Table~\ref{tab:DM2BIM}). In this framework, a graph-based method is adopted. Building upon the 1LSB results, containment relationships between SPACE and VOID entities, as well as adjacency between VOID entities, are evaluated. This enables flexible representation of complex geometries: for example, an atrium can be modelled as multiple vertically connected VOID entities bounded by a floor and ceiling, while a courtyard is represented as one or more VOID entities without an upper boundary.

\begin{figure}
	\centerline{\includegraphics[width=\columnwidth]{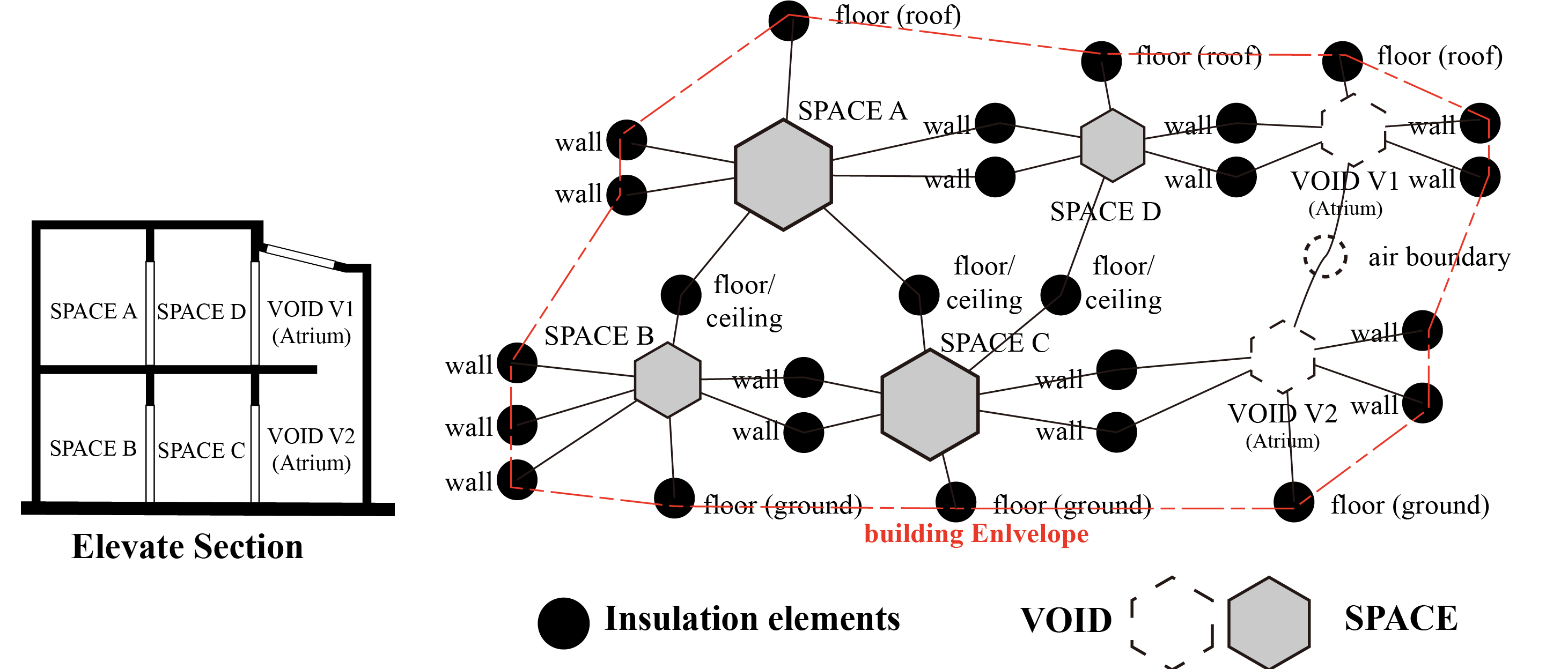}}
	\caption{space network as a graph composed by SPACE and VOID}
	\label{SpaceNetwork}
\end{figure}

Once the space-face associations are established, identifying space-space connectivity becomes straightforward, leading to the derivation of 2LSB relationships. The orientation factor for 2LSB is determined by evaluating the clockwise direction of each 2D 1LSB boundary. The final output is a connected thermal-zone network as illustrated in Fig. \ref{SpaceNetwork}.

To align the data requirement of the Ontologies, especially the Building Topology Ontology (BOT)\cite{Rasmussen2025} to record space information, the 1LSB and 2LSB data should be restructured hierarchically into the series of Model-Building-Storey-Zone-Element-Sub element. Besides, the 2LSB relations and the geometry vertices information are integrated into the structure as KG. 

\subsection{Building Information Model to Building Energy Model Transformation}
\subsubsection{Ontology Definition}
\begin{figure}
	\centerline{\includegraphics[width=\columnwidth]{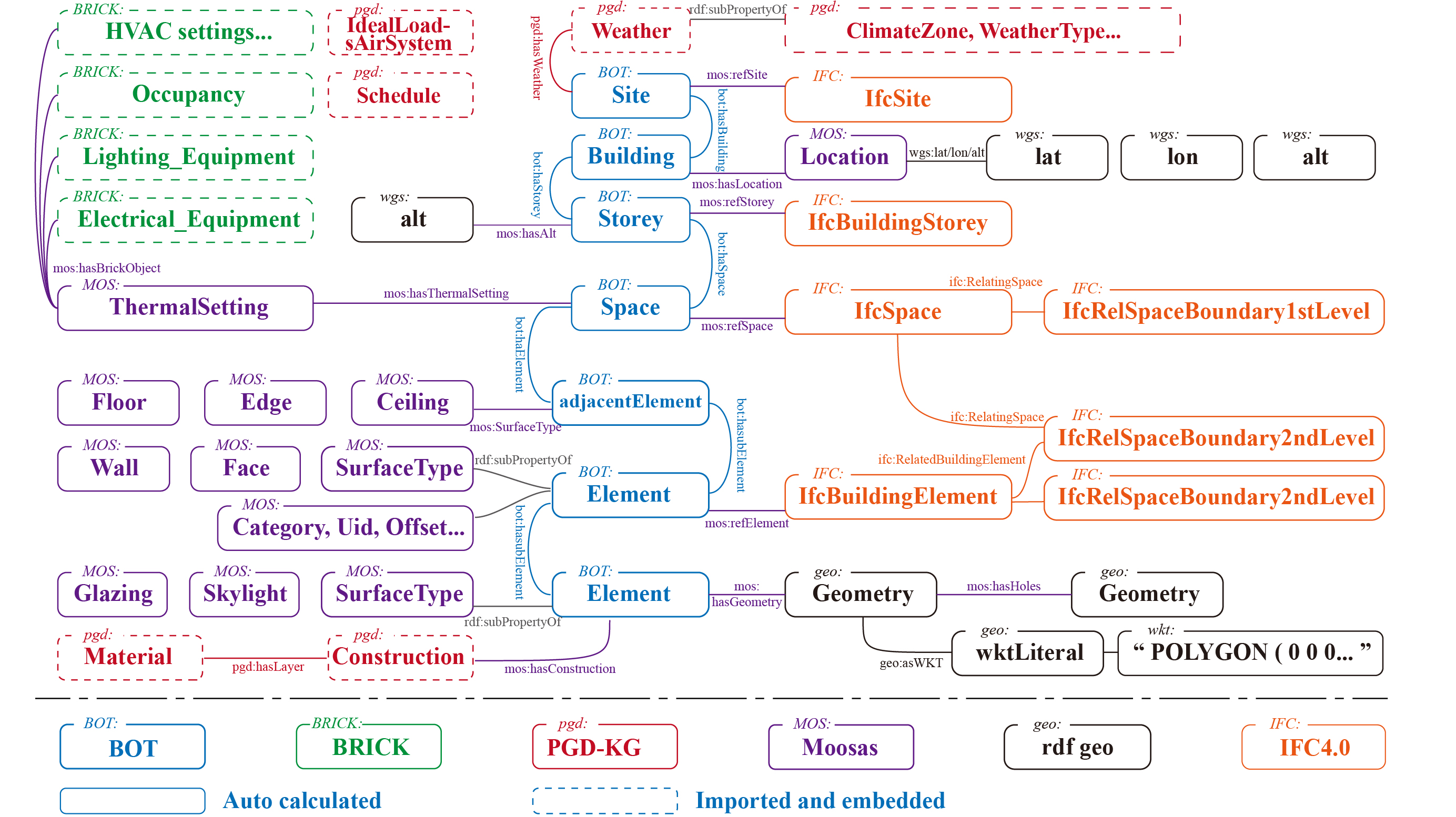}}
	\caption{Ontology Definition}
	\label{Ontology}
\end{figure}

Given the limited thermal information typically available in early-stage design models, external boundary conditions and thermal settings must be imported and integrated into the OWL model. To ensure consistent alignment of space topology, geometry, and thermal properties, several established ontologies are adopted as the foundation of the proposed model (Fig. \ref{Ontology}).

The ontology extends the KG-based BEM transformation schema (PGD) \cite{Wu2024knowledge} and incorporates geometry information from WGS and GEO \cite{World2024}, as well as 1LSB and 2LSB definitions from IFC 4.0 \cite{buildingSMART2025ifcOWL}, providing interoperability with CAD and IFC-based BIM. Basic HVAC and internal-gain concepts are integrated through the Brick Schema \cite{Brick2025}, while building topology follows the BOT ontology \cite{Rasmussen2025}. Components not covered by these existing ontologies (Brick, BOT, IFC4.0, WGS, PGD) are defined with the mos prefix for modularity and reproducibility.

The framework supports full export and import of KG-based model files (*.ttl, *.owl, *.xml). An additional alignment module enables export to *.gbXML, though *.gbXML import is not supported. Existing alignment tools (ifcOWL, Brick, DERIROOMS, etc.) can also be applied. However, due to the incomplete thermal data characteristic of early design, the alignment workflow by Wu et al.\cite{Wu2024knowledge} is not adopted; instead, a more flexible transformation method tailored to early-stage models is implemented.

\subsubsection{Data alignment between KG and IDF}
\begin{figure}
	\centerline{\includegraphics[width=\columnwidth]{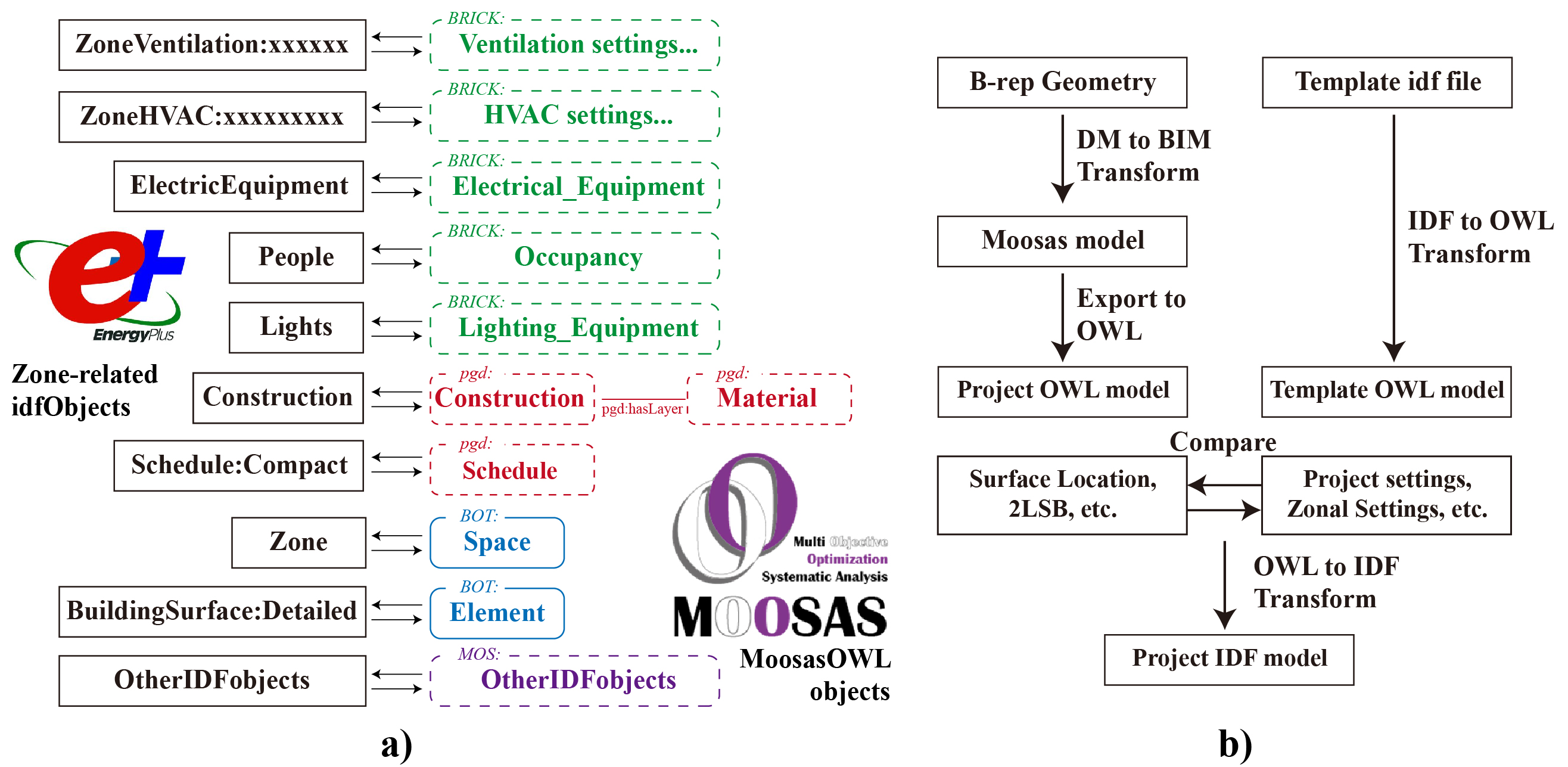}}
	\caption{a)Reversible transformation between BEM and KG b)The transformation pipeline.}
	\label{OntologyTransformation}
\end{figure}

Following existing transformation workflows, a bidirectional conversion module between KG and EnergyPlus *.idf is developed (Fig. \ref{OntologyTransformation}). Users may attach an *.idf template containing HVAC, construction, schedule, and activity parameters, which are then adapted and embedded into the KG model (Fig. \ref{OntologyTransformation}). During template processing, geometry and topology in the template file are ignored; only zone-level thermal attributes (HVAC, ventilation, internal gains) are preserved and mapped to all valid spaces derived from the B-rep model.

To ensure EnergyPlus compatibility, a standardization module is included. If BEM export is required, all vertical faces are reconstructed using the bottom-top 2D 1LSB projections, forming closed loops that guarantee full connectivity with floors and ceilings. This strictly satisfies EnergyPlus geometric constraints and preserves radiative heat-transfer accuracy. Users may opt to skip this reconstruction; in that case, EnergyPlus automatically falls back to a simplified solar distribution mode (FullExterior instead of FullInteriorAndExterior).

\subsection{Validation}
\begin{figure}
	\centerline{\includegraphics[width=\columnwidth]{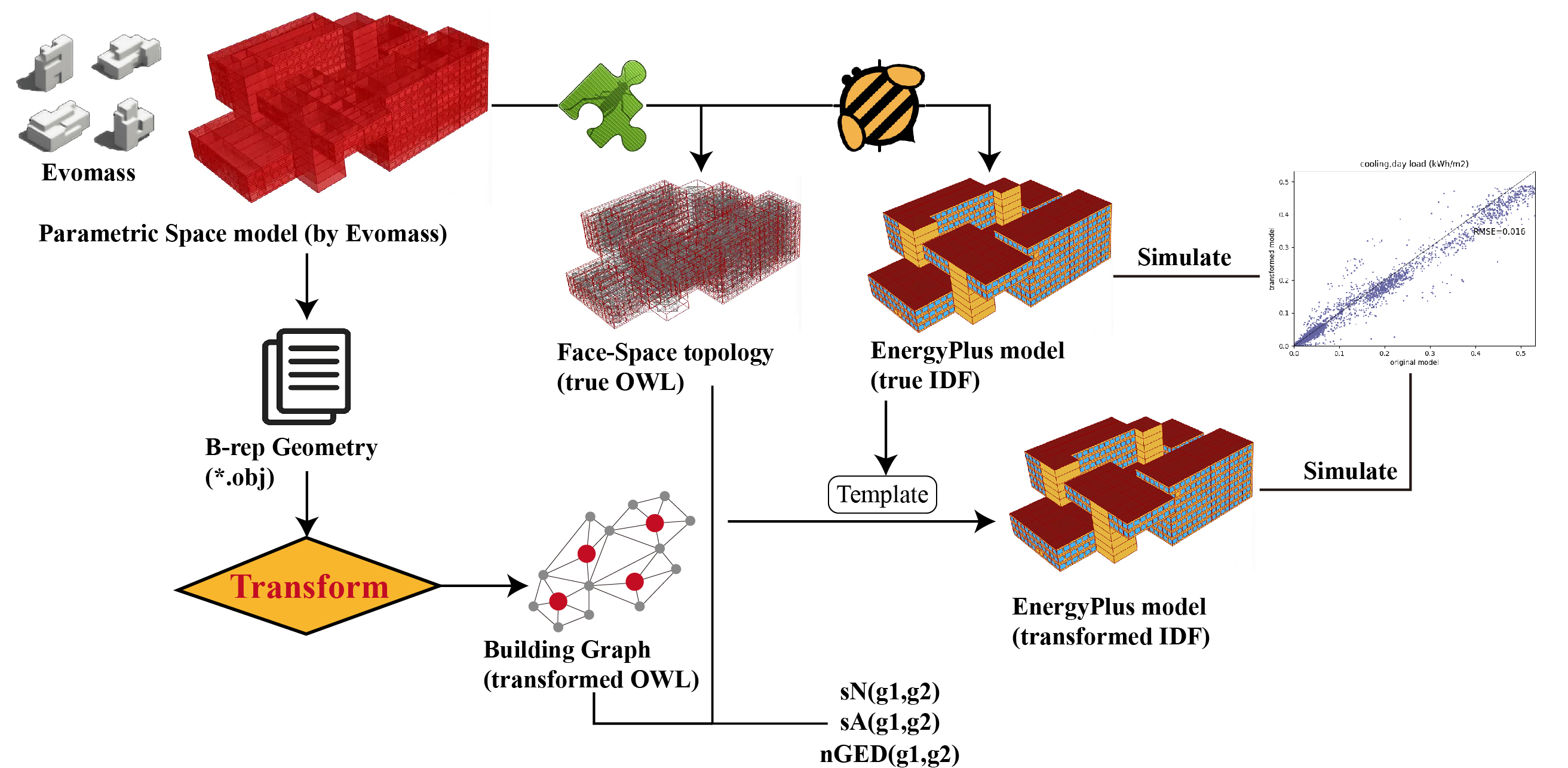}}
	\caption{Method for the internal performance testing based on generative datasets 1}
	\label{Dataset1Usage}
\end{figure}

The validation in this article focuses on both internal performance and comparison to existing solution. The validation 1) Topology accuracy and 2) Energy Model fidelity are targeted on the internal performance based on a experimental, generative dataset \ref{Dataset1Usage}. While the validation 3) Transformation Robustness Improvement focuses on the comparison and improvement to other raw ASG method on manually sketch model dataset. And finally the 4) Workflow Efficiency and Robustness focuses to the real applications and real buildings on SketchUp compared to commercial software SEFAIRA.

\begin{figure}
	\centerline{\includegraphics[width=\columnwidth]{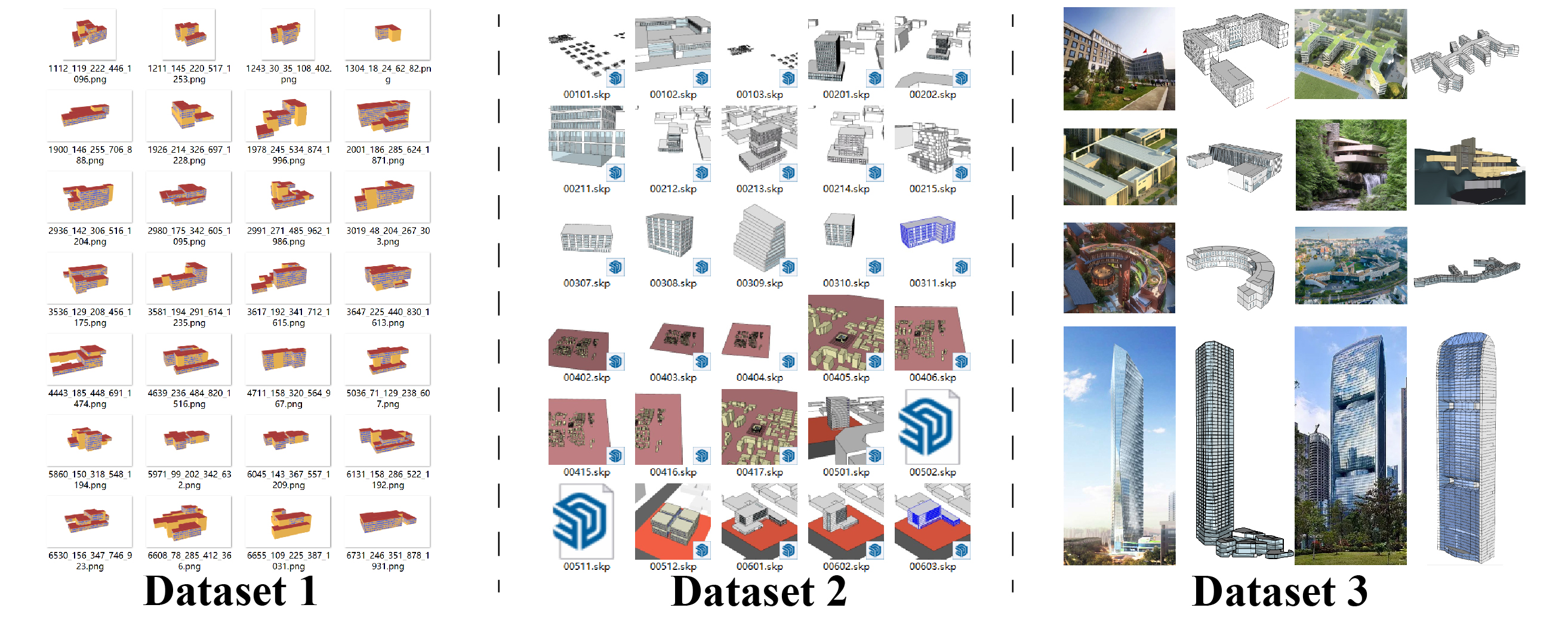}}
	\caption{The three datasets used in the validation}
	\label{dataset}
\end{figure}

This article introduces a large mixture B-rep model dataset (Fig. \ref{dataset}) 
to evaluate and refine the developed module. Comprehensive building models including 200 parametric models (dataset 1), 570 sketch design models (dataset 2), and 15 complex real-building models (dataset 3) was curated, for the validation on the topology accuracy, analytical model accuracy, robustness, the and the efficiency in different applications. 

A series of parametric models (dataset 1) are created by an auto-mass-generation module Evomass \cite{Wang2020} for the sake of derivable space topology and accurate *.idf modelling. The grasshopper module Evomass can create randomized multi-storey models with several shape parameters. A post-process script has been developed and applied to randomly create inner walls and windows to finish the thermal zoning on the parametric models following the grid axis of the building. The space volume would be maintained during the generation, which can be used to restore the space topology as well as build the energy model by Honeybee-grasshopper. Therefore, the dataset 1 is a completed auto-generated dataset with *.obj, ground truth *.owl, and ground truth *.idf for each case.

The sketch models (dataset 2) are sourced from architecture students and were specifically intended to emulate the characteristics of models produced during the early design stages. Accordingly, no restrictions are imposed regarding model cleansing, modeling techniques, or the specific design software utilized (provided an .obj file export was possible). To further challenge the module's robustness, several redundant or complex elements were intentionally introduced into the model testing sets. These included items such as shading devices, individual stair steps, and decorative façade components. The sole modeling constraint mandate that walls be represented as single surfaces rather than as two faces defining a thickness. Consequently, this dataset can be considered representative of the diverse range of model inputs encountered in practical scenarios. 

The real-building models (dataset 3) are collected from important green building design cases with relatively more complex spaces and geometries. The buildings were selected according to their complexity and shapes: Oblique and Curve shape contains non-orthogonal walls and faces creating small deviation and creak; Complex floor arrangement includes high space cross multiple storeys; High-rise Oblique shape have numeric faces and spaces requires high effort on calculation; and the Complex decorated building includes numeric redundant shading and decorating elements disturbing the space and face identification. All buildings are carefully modelled with their shape, storey, and façade. Especially, considering the applications in performance analysis, the connectivity between rooms with inner walls, doors and windows are restored. This dataset is included to validate the robustness of the framework’s application on real construction and design workflow.

\subsubsection{Topology accuracy}
The topology represented in the OWL models is validated using Dataset 1. A Grasshopper script is developed to reconstruct the space-face topology by extruding the restored space volumes. From the OWL models, a LPG is then generated, in which construction elements (walls, floors, glazing, skylights, etc.) and spaces serve as nodes, while space-face connections and wall-window relationships are represented as edges (Fig. \ref{SpaceNetwork}). 
Four evaluation metrics are designed and computed based on this graph structure.

Direct coupling between the generated graph and the ground-truth graph cannot be fully recovered during the design-model-to-BIM transformation, because the B-rep model retains only face-boundary information and excludes explicit semantic or topological labels. To address this, the ISOrank graph-alignment algorithm is employed to evaluate the correspondence between the two graphs. ISOrank computes a node-to-node Similarity score $(S_{}^{ab})$, which quantifies the structural alignment between the generated graph and the ground-truth topology. 

\begin{equation}
	\scalebox{0.85}{
		$S_{0}^{ab} = 
		\begin{cases} 
			\textstyle\min\left( \frac{D(a,b)}{0.5}, 1 \right) 
			\times \textstyle\min\left( \frac{Ar_a}{Ar_b}, 1 \right) 
			& \forall \, \mathrm{Type}_a = \mathrm{Type}_b \\
			0.0 
			& \forall \, \mathrm{Type}_a \neq \mathrm{Type}_b 
		\end{cases}$
	}
\end{equation}

\begin{equation}
	S_{i}^{ab} = \dfrac{S_{i-1}^{ab}}{N_{g_a} \cdot N_{g_b}} \cdot
	\sum_{p \in g_a} \sum_{q \in g_b} 
	\left[ (1 - \alpha) + \alpha \cdot S_{i-1}^{pq} \right]
\end{equation}

\begin{equation}
	N_{g_a} = \sum_{g_a} n,\quad N_{g_b} = \sum_{g_b} n
\end{equation}
Here $S_{0}^{ab}$ is the initial Similarity on a specific node twin $(a,b)$, one from the generated graph and the other from the ground truth graph;$S_{i}^{ab}$ is the Similarity in the iteration $i$;$D(a,b)$ is the spatial distance between $a,b$;$Ar$ is the spaces or faces area of the node;$N_{Ga}$ is the total number of nodes in the graph; and $\alpha$ is the control parameters for better conversion on the Similarity. The nodes could therefore be coupling by iterating the similarity matrix on all nodes on the two graphs.

Besides, a metric $s_{N}(g_a,g_b)$ indicating the total Similarity on the graphs nodes is introduced by \eqref{eq3}:
\begin{equation}\label{eq3}
	s_N(g_a, g_b) = \sum_{n \in g_a} \max_{n' \in g_b} \left\{ s^{(n, n')} \right\} / N_{g_a}
\end{equation}
To further illustrating the accuracy on space recognition, the Similarities on all space nodes is calculated by \eqref{eq4}: ($ns$ refers to the space nodes)
\begin{equation}\label{eq4}
	s_{NA}(g_a, g_b) = \dfrac{
		\sum_{n_s \in g_a} \left( Ar_{n_s} \cdot \max_{n_s' \in g_b} \left\{ S^{n_s, n_s'} \right\} \right)
	}{
		\sum_{n_s \in g_a} Ar_{n_s}
	}
\end{equation}
Thirdly, a traditional metric to illustrate the deviation during the generation is introduced \cite{Liu2025}. The Graph Edit Distance (GED) representing how many times we need to edit the graph Ga into Gb by adding, removing, or editing a node or edge. And the negative GED \eqref{eq6} is a standardized metric representing the accuracy:
\begin{equation}
	\text{GED}(g_a, g_b) = \min_{e \in \varepsilon} \sum c(e_i)
\end{equation}
\begin{equation}\label{eq6}
	\text{nGED}(g_a, g_b) = 1 - \dfrac{\text{GED}(g_a, g_b)}{\sum n + \sum e}
\end{equation}
Finally, a simple metric representing the calculation efficiency is introduced by \eqref{eq7}
\begin{equation}\label{eq7}
	E(g_a) = \dfrac{T}{\sum n_s}
\end{equation}

\subsubsection{Energy model fidelity}
The accuracy between the BIM and BEM is validated by the dataset 1. The same *.idf template is used in the generation of Honeybee BEM and the transformed BEM to ensure the same thermal settings. Any topology and geometric-related properties would not be preserved from the template to the transformed BEM including the external air connection or ground connection. Both the heating and cooling energy on multiple temporal resolutions are compared by \eqref{eq8} and \eqref{eq9}.
\begin{equation}
	\label{eq8}
	\text{RMSE}_m = \sqrt{
		\sum_{t}^{N_t} \frac{\left( E_t - \hat{E}_t \right)^2}{N_t}
	}
\end{equation}
\begin{equation}
	\label{eq9}
	R_t^2 = 1 - \dfrac{
		\sum_{t=0}^{N_t} \left( E_t - \hat{E}_t \right)^2
	}{
		\sum_{t=0}^{N_t} \left( E_t - \overline{E_t} \right)^2
	}
\end{equation}

Here the $E$ is the accumulated distinct cooling or heating energy for different temporal resolution in $kWh/m^2$.  Considering the sensitive period for cooling and heating, only the energy load on cooling/heating design days are recorded (288 timesteps for each case).

\subsubsection{Transformation robustness improvement}
This framework has included multiple methods to ensure the robustness including the complex cleanse module and the fuzzy matching on the element graph during the ASG process. Besides, the whole model would be checked and any problematic elements would be remodeled in the transformation BIM to BEM. To illustrate the contribution on robustness of these modules for the whole transformation process compared to original ASG methods, a robustness testing on dataset 2 with different combination of strategies is processed. Three metrics \eqref{eq7},\eqref{eq10}, and \eqref{eq11} are collected in the test: 
\begin{equation}
	\label{eq10}
	\text{sC} = \dfrac{\sum C_{\text{success}}}{\sum C}
\end{equation}
\begin{equation}
	\label{eq11}
	\text{sA} = \dfrac{\sum Ar_{n_s}}{GA}
\end{equation}
$sC$: the successfully transformed cases / the total cases.
$sA$: the recognized total floor area (${Ar}_{ns}$) / record gross floor area ($GFA$)

The performance of the individual cleansing methods is evaluated, followed by a comparison of two application scenarios to the original ASG method:

\underline{Automatic transform.}
DM-to-BEM transformation is executed directly on raw design models without any pre-processing. This workflow is targeted on the fully-automatic pipeline, especially those with AI-control processing. Although computationally efficient for simple geometries, this workflow is not robust and frequently fails when encountering illegal or malformed faces.

\underline{Semi-automatic transform.}
DM-to-BEM transformation is applied to models manually cleaned by users. This workflow is targeted on the human-driven interactive pipeline which always embedded in the modeling tools. In practice, the Intersect Faces with Model tool in SketchUp is executed, and decorative or non-thermal elements (e.g., stairs, shading devices, columns, beams) are removed. This improves face identifiability and reduces topological ambiguity.

\subsubsection{Workflow efficiency and robustness}
\begin{figure}
	\centerline{\includegraphics[width=\columnwidth]{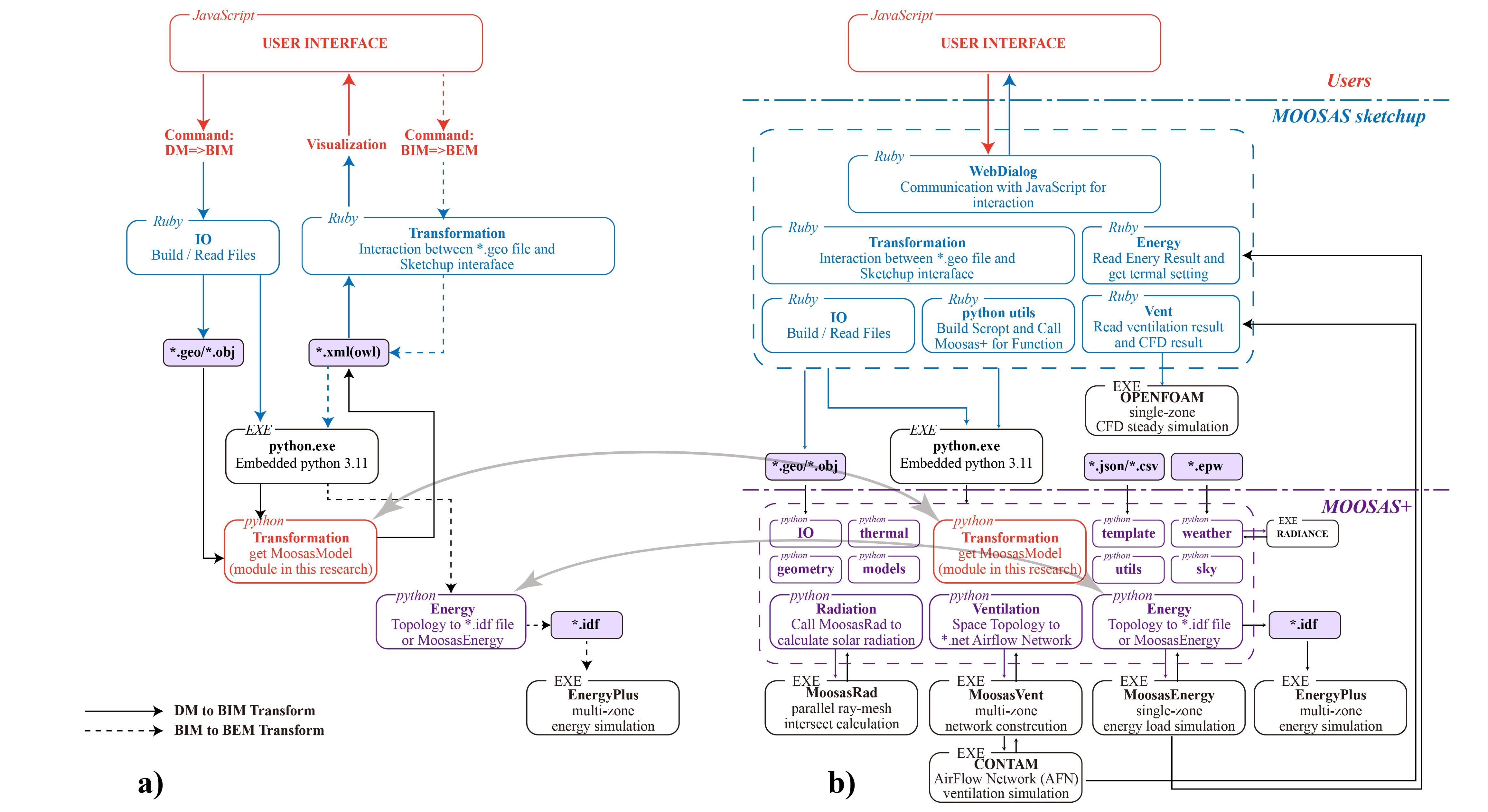}}
	\caption{The embedding of the framework into Moosas+SketchUp. a) The transformation workflow inside the embedded framework. b) The whole Moosas+SketchUp Software Framework.}
	\label{MoosasPlus}
\end{figure}

This validation focuses on assessing the framework in a realistic design workflow. A SketchUp-based interface has been developed (Fig. \ref{MoosasPlus}), enabling users to export B-rep data, invoke the transformation pipeline, import the generated ontology model (.xml), and produce the corresponding EnergyPlus model (.idf). To improve I/O performance, the interface employs a custom stream-based format (*.geo). This integration allows the framework to be embedded within the open-access simulation platform MOOSAS \cite{Lin2021MOOSAS} and extended into a complete building simulation environment.

For benchmarking, the commercial software SEFAIRA \cite{Sefaira2023} is evaluated using Dataset 3. Because SEFAIRA does not provide an accessible API or automated workflow, all transformations are performed manually and documented. Efficiency is assessed using the metric in Eq. 7, with processing time captured via a mouse-tracking tool. Robustness is evaluated using Eq. 11. Each case is executed multiple times to minimize the influence of manual operations and other stochastic factors.

\section{Result}
\subsection{Internal Performance: Topology Accuracy}
\begin{figure}
	\centerline{\includegraphics[width=\columnwidth]{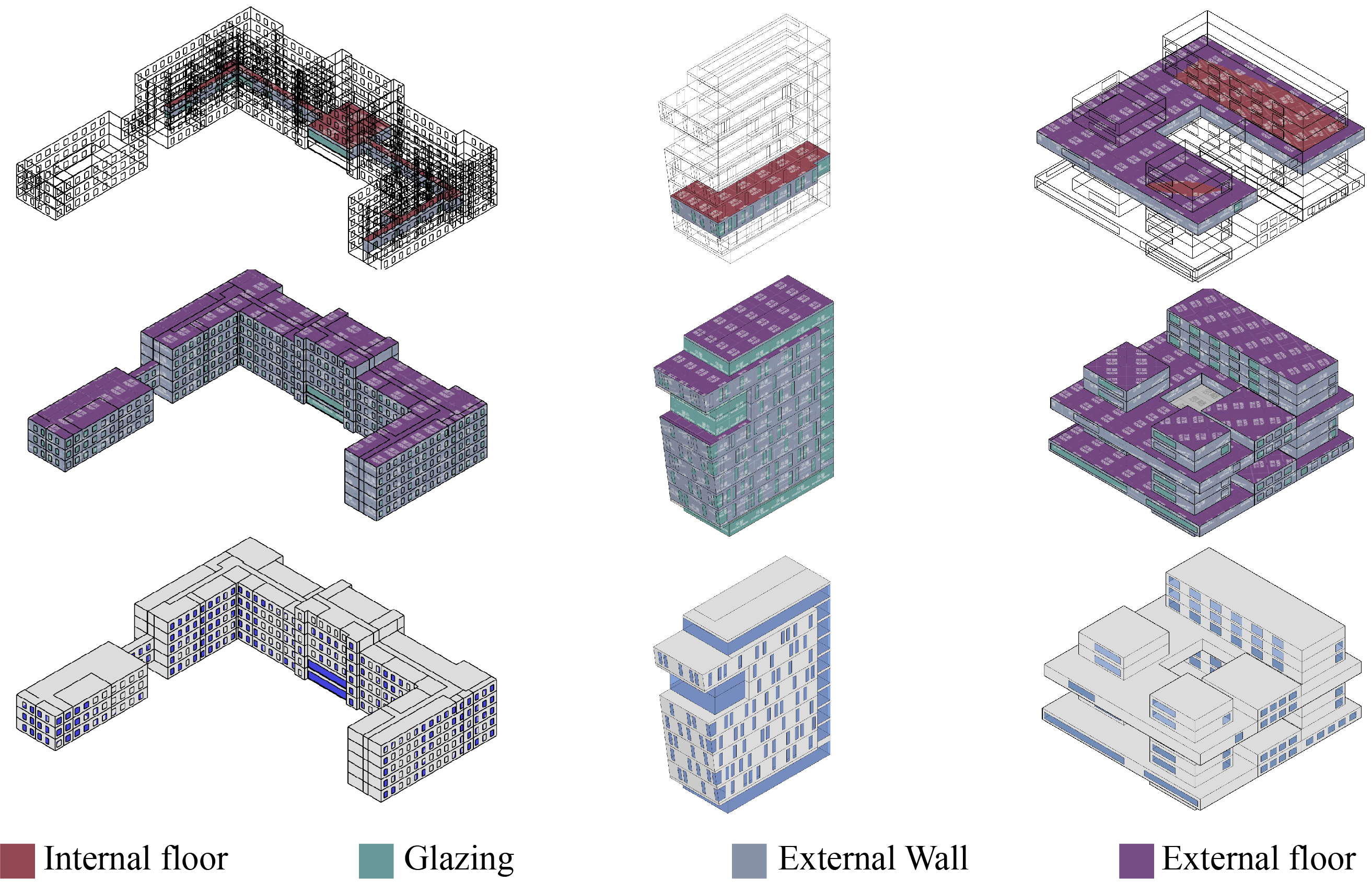}}
	\caption{Transformation Visualization by the MOOSAS-SketchUp interface}
	\label{Visualization}
\end{figure}
\begin{figure}
	\centerline{\includegraphics[width=\columnwidth]{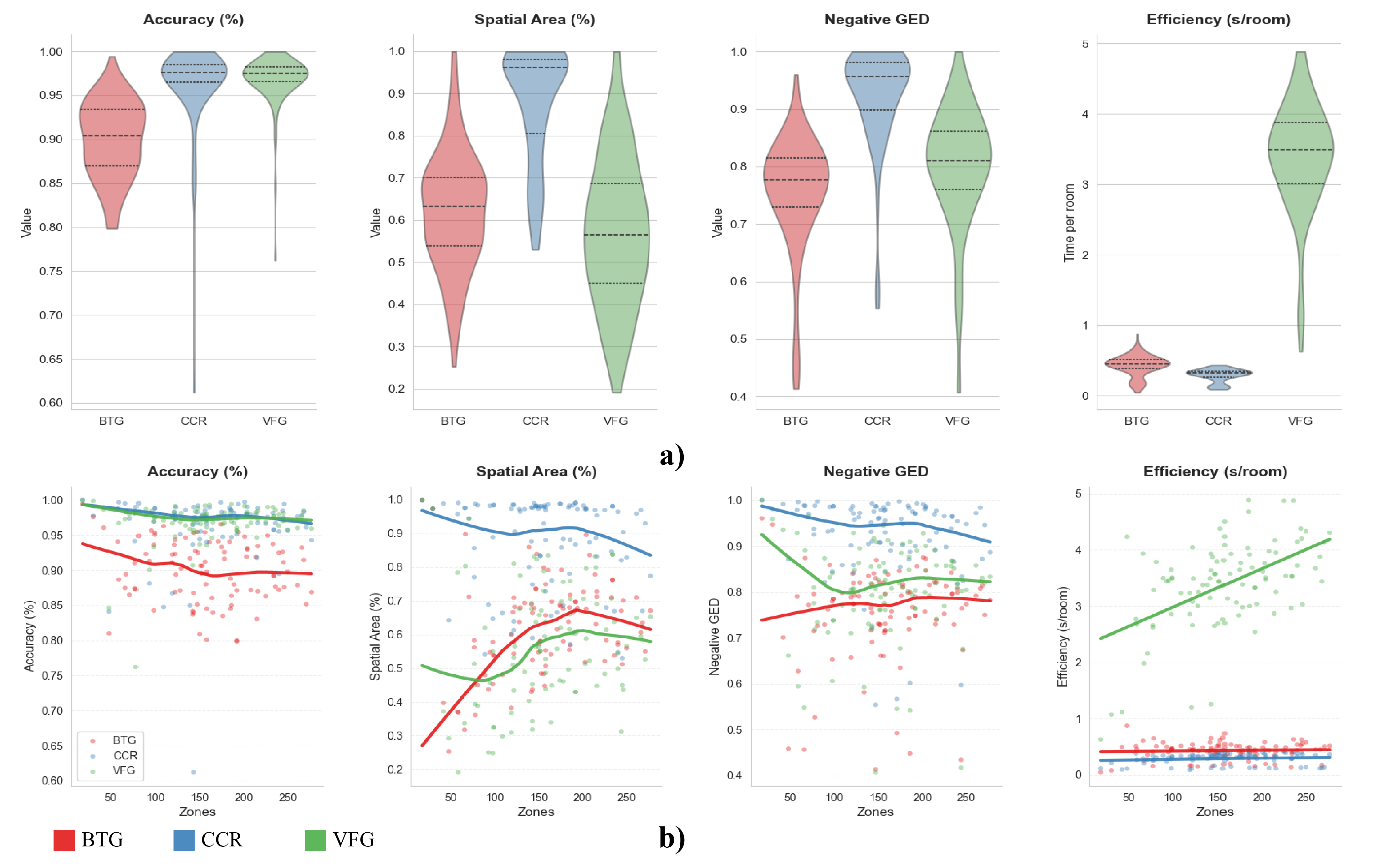}}
	\caption{KG Topology validation result on datasets 1. a) Distribution of the metric on the three methods b) performance changes with the number of zones on the testing cases.  Locally Weighted Scatterplot Smoothing (LOWESS) method is used to construct the fitting lines.}
	\label{topoResult}
\end{figure}
Fig. \ref{Visualization} illustrates examples of the transformation function applied to three distinct cases from the dataset. And 
Fig. \ref{topoResult} presents the topology accuracy and computational efficiency of the three embedded ASG methods—BTG, CCR, and VFG—used to transform design models (DM) into OWL-based BIMs. Overall, the CCR method shows the strongest balance of accuracy and performance, achieving an average topological similarity of ${nGED}(g_a, g_b) = 0.87$ and an efficiency of $E(g_a) = 0.290;\text{s/room}$.

For element identification accuracy (${s_N}(g_a, g_b)$), VFG exhibits clear advantages. Its visibility-based surface connectivity enables the highest average accuracy (0.968), outperforming CCR (0.962) and BTG (0.899). BTG struggles in irregular spaces where walls fail to connect cleanly with ground and ceiling surfaces—such as bedrooms with overhanging window ledges. These cases are reliably identified by VFG, whereas CCR partially captures them but may omit certain horizontal faces that require reattachment in later processing. As room count increases, all three methods show declining ${s_N}(g_a, g_b)$ performance.

Unlike ${s_N}$, the metric ${s_{NA}}(g_a, g_b)$ evaluates both the number of detected space elements and their 2D boundary shapes. Here, CCR performs best, achieving an average ${s_{NA}}$ of 0.875, compared with 0.570 for VFG and 0.620 for BTG. VFG’s low ${s_{NA}}$ indicates that although it reliably locates space elements, it often fails to assemble them into coherent, closed boundaries. This limitation is correctable by tightening connectivity heuristics, but such improvements require substantially more topological computation. VFG and BTG also exhibit instability in ${s_{NA}}$ across models, especially those with only a few thermal zones.

Topological similarity, captured by ${nGED}(g_a, g_b)$, provides the most holistic indicator of graph-level correctness—a critical factor for downstream KG-based BEM generation. CCR again performs best (0.871), followed by VFG (0.753) and BTG (0.710). The variability in VFG stems from its stochastic ray-generation process. CCR’s performance also degrades with increased model complexity, reducing both ${s_{NA}}$ and ${nGED}$.

CCR further delivers the highest computational efficiency. The metric $E(g_a)$ incorporates cleansing, post-processing, and 2LSB extraction (Table \ref{tab:topoEff}). BTG exhibits the highest computational cost because it must evaluate full face-face connectivity for both vertical and horizontal surfaces (Fig. \ref{ASG}), then reconstruct wall loops—operations involving repeated Boolean calculations. BTG also inserts air-wall elements when boundaries are incomplete, increasing adjacency-solving overhead relative to CCR and VFG. In VFG, robustness improves with more rays, but this causes an exponential increase in ray-face intersection checks, making VFG substantially less efficient.

On the consideration of complexity, (Table \ref{tab:topoEff}) most of the physical calculations based on set operation were claimed as $O(n^2log n)$ based on the $O(nlog n)$ for face intersection operation. Two layers of traversal including an intersection operation inside are required in VFG method which leads to the lowest efficiency for this 1LSB process.

\begin{table}[htbp]
	\centering
	\caption{Detail composition of the calculation duration for the three methods on datasets 1.}
	\label{tab:topoEff}
	{  
		\begin{tabular}{l c c c c c c c c}
			\toprule
			\multirow{2}{*}{s/room}  & \multirow{2}{*}{I/O} & \multirow{2}{*}{Face Class.} & \multirow{2}{*}{Cleansing} & \multicolumn{3}{c}{1LSB} & \multirow{2}{*}{Space Pack.} & \multirow{2}{*}{2LSB} \\
			\cmidrule(lr){5-7} 
			&&&&BTG&CCR&VFG& & \\
			\midrule
			Duration & 0.001 & 0.138 & 0.010 & 0.088& 0.012 & 3.103 & 0.125 & 0.025 \\
			\midrule
			Complexity & $O(n)$ & $O(n^2log n)$ & $O(n^2log n)$ & $O(n^2)$ & $O(n log n)$ & $O(n^3log n)$ & $O(n^2log n)$ & $O(n^2)$ \\
			\bottomrule
		\end{tabular}
		\\[1ex] 
		\raggedright\footnotesize *The complexity was briefly estimated based on the algorithm. The input scale is different in each process.
	}
\end{table}

\subsection{Internal Performance: Energy Model Fidelity}
\begin{figure}
	\centerline{\includegraphics[width=\columnwidth]{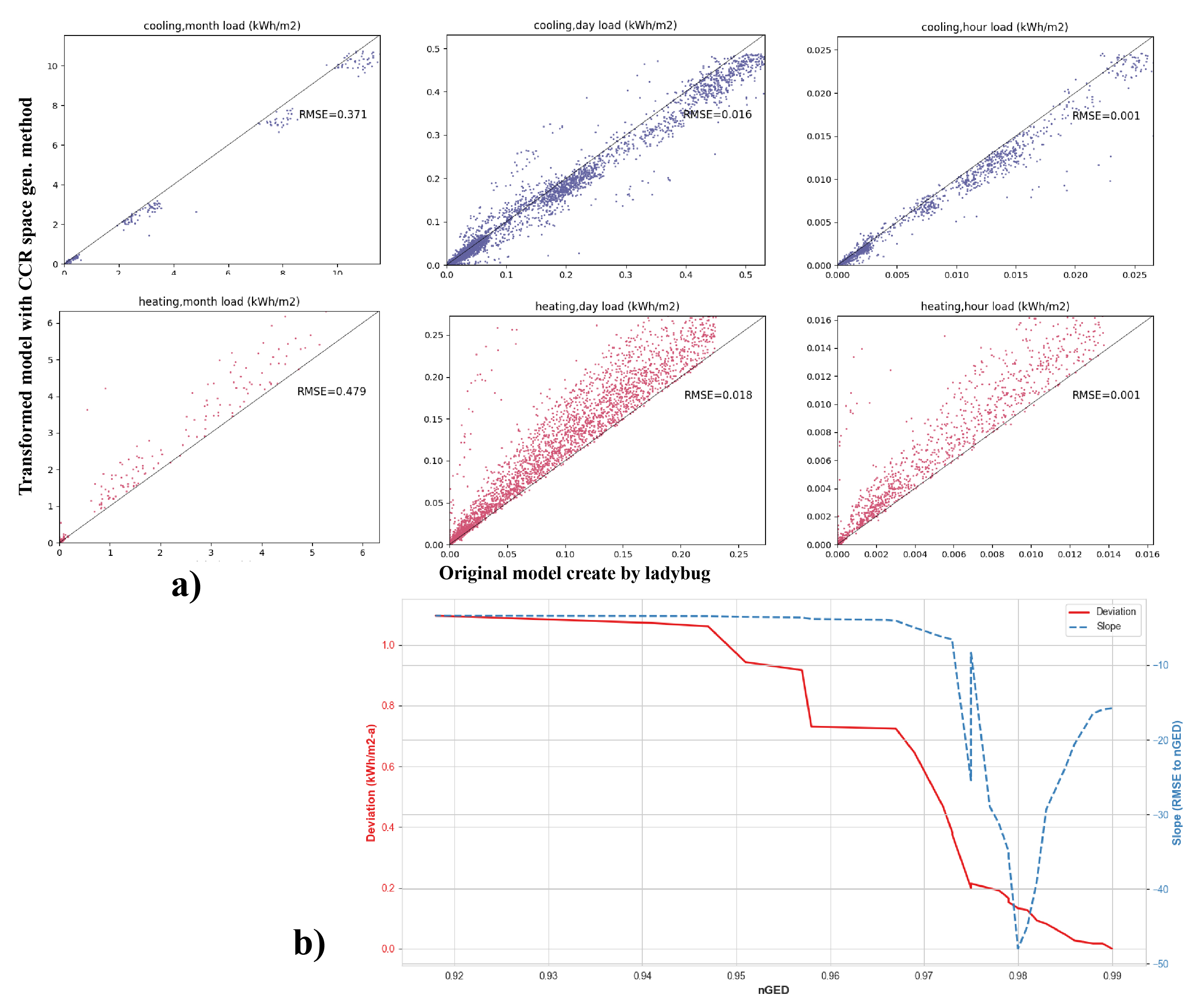}}
	\caption{Energy model fidelity for the whole transformation from DM to BEM tested on datasets 1. a) simulation accuracy b) sensitivity of deviation and nGED.}
	\label{BEMAcc}
\end{figure}

Fig. \ref{BEMAcc} presents the energy-simulation accuracy. For typical usage scenarios, the full DM-to-BEM workflow using the CCR-based 1LSB method is evaluated, meaning that topology errors introduced during DM→BIM propagate into the BIM→BEM stage.

The robustness of template-setting embedding is reflected in the lighting and appliance results, which show near-zero RMSE and $R^2 \approx 1.0$ (in which $RMSE<10^{-12},\text{kWh/m}^2$, attributable to EnergyPlus numerical noise). Heating and cooling loads exhibit small deviations from the ground-truth Honeybee *.idf models, achieving monthly $R^2$ of 0.946 and 0.987 with RMSE of 0.51 and 0.50 $kWh/m^2$. The high $R^2$ reveals two advantages in this module. Firstly, the validation here strictly measures the geometric fidelity passed to EnergyPlus, which means this module have high geometric fidelity in the transformation, although a standardization was implemented during the BIM→BEM stage; Secondly, the thermal settings are populated via the same IDF template to the ground truth, which means this module also have well settings fidelity.

Except for the accumulated error from topology, remaining discrepancies arise from two sources:
(1) Ground-connection inconsistencies. Ground-floor detection differs between Ladybug tools and this framework: all lowest-level floor planes are treated as ground surfaces here, due to current data-structure limitations.
(2) Face standardization. To ensure robustness, all faces are reconstructed from their 2D projections before *.idf generation, slightly altering geometries and introducing minor simulation deviations.

The sensitivity analysis on $nGED$ and annual $RMSE$ (Fig. \ref{BEMAcc}-b) reveal the impact from low model fidelity on energy simulation accuracy. The energy simulation is incorrect with $nGED$ lower than 0.9 which have obvious mistake on building shape. Users should priorly fix the model until there is no building shape misinterpretation. For those with $nGED > 0.9$, very small differentiation for cases with $nGED \approx 0.98$ will have significant changes in energy intensity accuracy. According to the record on $sA$, the cases with $nGED > 0.98$ would only have topology error in the internal partitions or apertures; while the cases with $nGED < 0.98$ usually have mistake on the building envelope and boundary. As a conclusion, the envelope is the most essential element to ensure the model fidelity on BEM, which should be prioritized in the transformation.

\begin{table}[htbp]
	\centering
	\caption{Summary of the metric on $s_{N}(g_a,g_b)$, $s_{NA}(g_a,g_b)$, ${nGED}(g_a,g_b)$ for CCR method on datasets 1.}
	\label{tab:CCRTopoAcc}
	{  
		\begin{tabular}{l c c c c c c c c}
			\toprule
			& \multicolumn{1}{c}{min} & \multicolumn{1}{c}{0.1} & \multicolumn{1}{c}{0.2} & \multicolumn{1}{c}{0.3} & \multicolumn{1}{c}{0.5} & \multicolumn{1}{c}{0.7} & \multicolumn{1}{c}{0.9} & \multicolumn{1}{c}{max} \\
			\midrule
			$s_{N}$     & 0.612 & 0.939 & 0.957 & 0.967 & 0.975 & 0.983 & 0.991 & 1.000 \\
			$s_{NA}$     & 0.530 & 0.652 & 0.754 & 0.831 & 0.957 & 0.977 & 0.987 & 0.999 \\
			$nGED$   & 0.001 & 0.674 & 0.858 & 0.897 & 0.948 & 0.977 & 0.988 & 1.000 \\
			\bottomrule
		\end{tabular}
	}
\end{table}

\subsection{Transformation Robustness Improvement on raw ASG}
\begin{figure}
	\centerline{\includegraphics[width=\columnwidth]{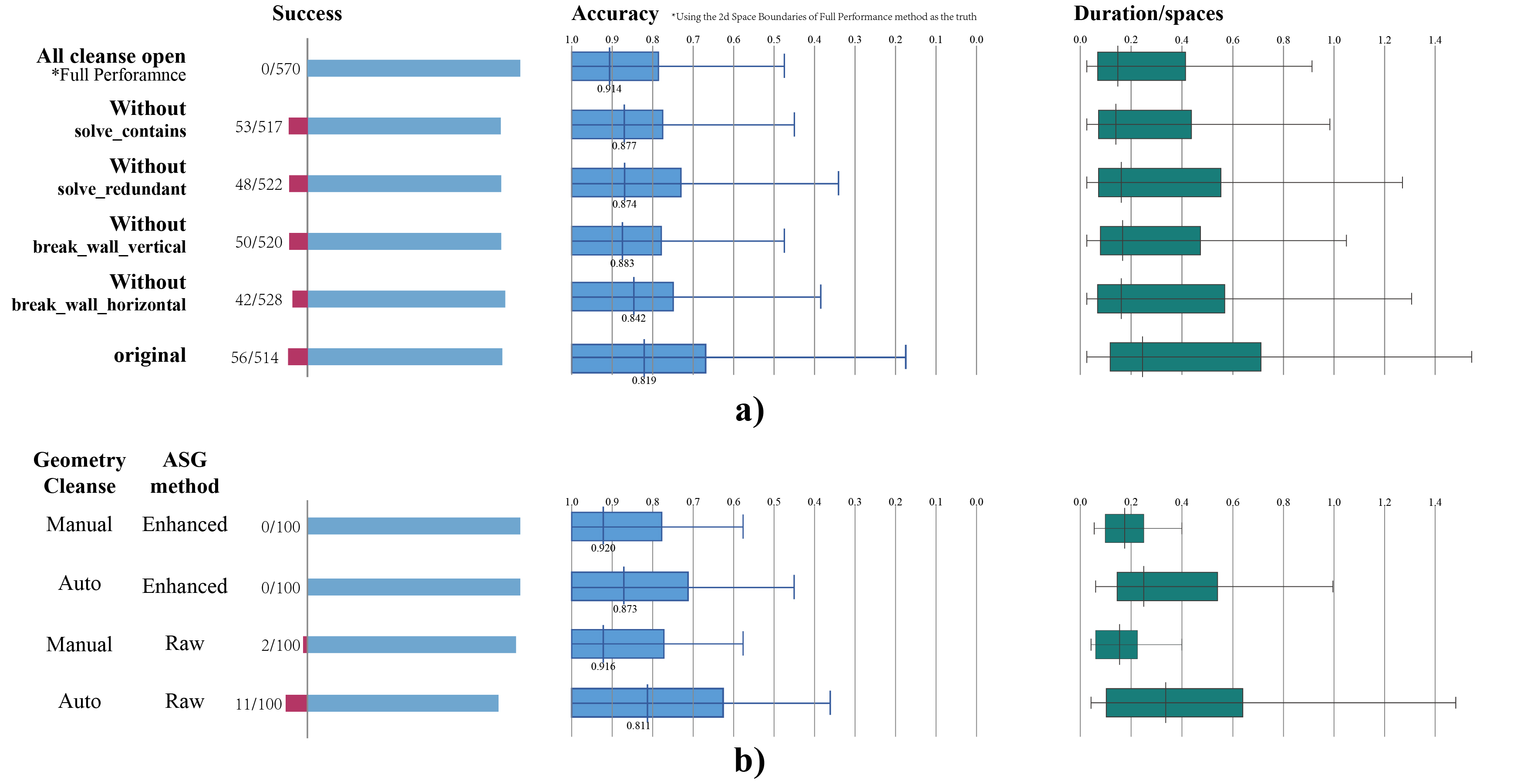}}
	\caption{Comparison between cleanse method tested on datasets 2. a) The implementation on four cleanse methods. b) Combination with the manual cleanse in SketchUp.}
	\label{Robustness}
\end{figure}

This validation assesses the robustness of the framework in early-design scenarios. Key performance metrics—robustness $sC$, spatial accuracy $sA$, and processing time $E(g_a)$—are summarized in Fig. \ref{Robustness}. Across all cleansing configurations, the cleansing module consistently improves both accuracy and robustness. Full automatic cleansing increases average robustness from 89\% to 100\% and spatial accuracy from 71.9\% to 81.4\%. It also improves efficiency by reducing problematic geometries encountered during topology generation.

A subset of 100 models from the 570-model dataset—characterized by complex forms, internal partitions, and interfering elements such as structural members and shading devices—was manually cleaned in SketchUp for comparison across two workflows. These models were carefully checked and edited with complex spatial and geometrical topology but higher quality on the faces (with less irregular faces and redundant components like stairs.)

The framework show improvement on the raw ASG method. It has improved the spatial accuracy $sA$ from 83.4\% to 87.1\% on the fully automatic pipeline. Besides, although the cleanse and pre-classification module has resulted in a computation time of at least 0.15 seconds/space, the enhanced geometries accelerates the calculation with less geometrical error and better robustness. It shows that without the manual cleanse or remodel effort, this framework can show obvious improvement in accuracy, robustness, and efficiency.

Besides, the result shows the limited substitutability of the cleanse module to manual processing. Automatic transform with original CCR method, which process the transformation without manually editing on geometries, show limited spatial accuracy compared to the semi-automatic workflow. In comparison, combining SketchUp’s intersection tools with manual removal of non-spatial elements, the Semi-automatic transform workflow with manual geometric processing substantially improves performance, failing only in two cases and raising $sA$ from 83.4\% to 91.6\%. Besides, the  This observation suggests that the manual processing have significant improvement on the performance; while the enhancement of the framework still could not fully take place the manual processing since the enhanced ASG method with automatic cleanse could only preserve $sA$ with 87.1\%.

\subsection{Workflow Efficiency and Robustness Compared to Existing Solution}
To evaluate the framework in real-world applications, it was embedded in MOOSAS and compared with the commercial software SEFAIRA on the SketchUp platform. As SEFAIRA is closed-source, face classification and energy analysis were executed and recorded manually, using accessible metrics $sA$ and $E(g)$ (with $GFA$ measured directly from the model) (Table~\ref{tab:software_comparison}). SEFAIRA does not provide spatial topology results; its interface only displays automatically classified face tags and performs BEM transformation and analysis as a single process. In this evaluation, face classification—including partial cleansing—is decoupled from the remainder of transformation in the framework, while MOOSAS Energy is used to assess the complete DM-to-BPS workflow. Besides, to exclude the I/O issues and focus on the main efficiency and accuracy, before the start of transformation, the *.geo files were exported from sketchUp models which will directly run the SEFAIRA transformation.

SEFAIRA exhibits more stable and efficient face classification in simple cases, but its performance declines for high-rise buildings (cases 12) due to delays from complex cleansing. In buildings with intricate spatial arrangements (cases 6), SEFAIRA shows poor robustness and fails to complete the full transformation. Notably, in case 15 (Fallingwater Apartment), which contains numerous shading, decorative, and structural elements, GFA deviation reaches 638\%, indicating failure to filter redundant items.

In contrast, the proposed framework maintains high stability across all cases, with a maximum $sA$ error of 22\%. Case 5, a curved building, exhibits $sA = 0.78$, highlighting limitations in handling irregular geometries. For case 15, GFA is overestimated because massive walls are misidentified as spaces, suggesting the need for additional filtering of invalid spaces in future versions.
\begin{table}[htbp]
	\centering
	\caption{Compare the application on real building to commercial software SEFAIRA on datasets 3.}
	\label{tab:software_comparison}
	{  
		\begin{tabular}{l c c c c c c c c}
			\toprule
			\multirow{3}{*}{case} & \multirow{3}{*}{GFA ($m^2$)} & \multicolumn{3}{c}{This Study} & \multicolumn{3}{c}{SEFAIRA} \\
			\cmidrule(lr){3-5} \cmidrule(lr){6-8}
			& & \multicolumn{2}{c}{E(g) (s)} & \multirow{2}{*}{sA} & \multicolumn{2}{c}{E(g) (s)} & \multirow{2}{*}{sA} \\
			\cmidrule(lr){3-4} \cmidrule(lr){6-7}
			& & Classify & Analyze & & Classify & Analyze & \\
			\midrule
			1 & 18617.05 & 8.06 & 12.72 & 1.00 & 2.54 & 107.11 & 1.21 \\
			2 & 390.80 & 0.61 & 2.00 & 1.00 & 2.70 & 39.74 & 2.15 \\
			3 & 83948.63 & 16.57 & 104.86 & 0.82 & 2.89 & 122.39 & 1.32 \\
			4 & 23247.22 & 2.96 & 25.16 & 0.86 & 2.55 & 73.03 & 1.26 \\
			5 & 4793.64 & 1.13 & 5.44 & 0.78 & 2.13 & 36.32 & 1.37 \\
			6 & 8128.79 & 6.34 & 6.23 & 1.02 & N/A & N/A & N/A \\
			7 & 447471.00 & 1.97 & 1.46 & 1.01 & N/A & N/A & N/A \\
			8 & 156403.00 & 2.73 & 7.38 & 0.96 & 2.75 & 122.83 & 1.79 \\
			9 & 261670.16 & 4.60 & 22.07 & 0.90 & 2.92 & 167.50 & 3.50 \\
			10 & 127673.28 & 18.78 & 201.59 & 1.00 & N/A & N/A & N/A \\
			11 & 25745.77 & 0.80 & 6.18 & 0.99 & N/A & N/A & N/A \\
			12 & 447206.65 & 18.16 & 64.77 & 1.00 & N/A & N/A & N/A \\
			13 & 196788.64 & 13.06 & 51.95 & 1.12 & N/A & N/A & N/A \\
			14 & 170383.69 & 11.98 & 40.21 & 1.03 & 2.28 & 212.71 & 0.93 \\
			15 & 395.69 & 3.27 & 5.88 & 1.20 & 2.58 & 252.84 & 7.38 \\
			\bottomrule
		\end{tabular}
	}
\end{table}

\section{Discussion}
This module has been incorporated into the MOOSAS+ simulation software, a standalone application developed using Python and Go. By integrating with the BPAM module existed in MOOSAS, this framework can actually not only provide the analytical model of Energyplus, but also for RADIANCE, COMTAM, and OPENFOAM, which all from a B-rep design model with maximized design flexibility. 

In conventional BPS workflows, the integration with grey-box BPS methods or surrogate models such as GNNs or KG-LLMs can fully automate and accelerate the simulation-optimization loop. On the general pictures of DM-BIM-BEM, there are four practical software/frameworks for this automatic pipeline (Table~\ref{tab:pipeline_comparision}). In comparison, this framework is the first open-access and fully automated tools fitting the applications on the AI intervention on DM-BIM-BEM. It also shows significant advantage in transformation robustness and accuracy (Table~\ref{tab:software_comparison}), providing more practical and stable function in the AI or BPS workflow. Besides, beyond EnergyPlus, the framework is fully compatible with MOOSAS Energy \cite{Lin2021MOOSAS}, a grey-box energy-load estimation tool, achieving analysis times of 0.01 s per zone and 0.29 s for the full DM-to-BEM transformation. This realizes the intervention of applications with higher efficiency requirement, like performance-orient optimizations or interactive BPS in the early design.

\begin{table}[htbp]
	\centering
	\caption{comparison on the level of automation to existing software providing the whole DM to BIM workflow}
	\label{tab:pipeline_comparision}
	{  
		\begin{tabular}{l c c c c}
			\toprule
			& \textbf{This study} & \textbf{SEFAIRA} & \textbf{Cove.tools} & \textbf{Ladybug} \\
			& Open-access & Closed & Closed & Open-access \\
			\cmidrule(lr){1-5}
			Cleansing & Auto & Semi-Auto & Semi-Auto & Manual \\
			Classification & Auto & Auto & Manual & Auto \\
			ASG (1LSB) & Auto & Auto & Auto & Manual \\
			Topology (2LSB) & Auto & Auto & Auto & Semi-Auto \\
			Thermal Embedding & Auto & Auto & Auto & Auto \\
			BPAM Generation & Auto & Auto & Auto & Auto \\
			\bottomrule
		\end{tabular}
	}
\end{table}

\subsection{Limitation}
Despite its robustness and automation, the framework exhibits several limitations. 

The cleansing module emphasizes transformation robustness over accuracy, yielding smaller gains in $sA$ than semi-automatic workflows, which may constrain fully automated applications such as building performance optimization or AI-based preprocessing (GNN/KG-LLM), though manual correction is straightforward. Efficiency is also limited, with face classification and space construction consuming on average 83\% of processing time, highlighting a key area for optimization. Geometric modifications during cleansing can introduce errors, including self-intersections from CSG operations (by pygeos), requiring future remediation. 

The Auto Space Generation (ASG) module remains one of the key technical challenges in the proposed pipeline and could be further extended through the incorporation of additional algorithms and systematic benchmarking. In particular, pixel-based approaches represent a promising direction that should be emphasized in future work. Such methods have demonstrated strong robustness in two-dimensional closed-contour detection (2D-1LSB), which is conceptually related to the iterative CCR-based topology solutions implemented in AutoCAD. Moreover, pixel-based approaches offer natural compatibility with computer-vision-based machine learning models, enabling the integration of visual learning techniques for spatial boundary detection \cite{CVPRFloorASG}. This integration may further enhance robustness when processing noisy or imperfect B-rep geometries. Nevertheless, within the current framework the 1LSB generation stage accounts for only a small fraction of the overall computational cost. Therefore, although pixel-based ASG methods show promising theoretical advantages, their practical benefits within the complete DM→BEM pipeline still require further empirical testing and comparative evaluation.

With further optimization of the geometry cleansing module and the inclusion of additional ASG strategies, the robustness of the framework when processing irregular buildings could be improved. In Dataset 3, Case 5 (a curved building) exhibits the lowest $sA$ value, indicating insufficient robustness when handling non-orthogonal geometries. Two primary barriers contribute to this limitation. First, the input B-rep models are typically polygonised , which introduces numerous small gaps and discontinuities between walls and floors. These artifacts significantly degrade the quality of the derived graph representation and reduce the effectiveness of topology-based ASG methods. Alternative approaches based on physical or geometric reconstruction may help improve the resulting $sA$. Second, polygonisation often separates opaque surfaces from their associated apertures. As a result, apertures may be incorrectly interpreted as curtain walls, which disrupts the overall topology identification process. Addressing this issue requires a more robust aperture-surface attachment mechanism during the face classification stage to ensure consistent semantic and topological relationships.

Finally, the framework does not yet implement automated subdivision of contiguous thermal zones or elongated spaces (e.g., corridors), which is essential for accurately capturing variable thermal loads and generating robust building energy models. Future development will target enhanced automatic cleansing, improved efficiency, geometric validation, and automated thermal zoning to support fully end-to-end BPS workflows.

\subsection{Applications}
\begin{figure}
	\centerline{\includegraphics[width=\columnwidth]{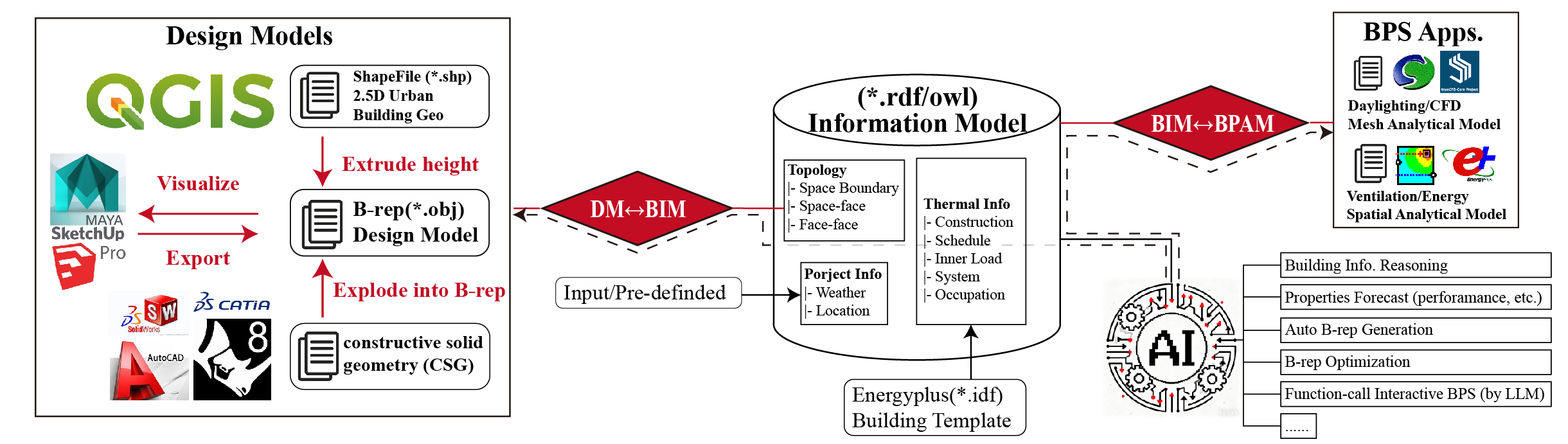}}
	\caption{Potential applications with this framework.}
	\label{Applications}
\end{figure}

This framework is the first to provide a transformation from unstructured B-rep models to graph-based formats—both LPG and KG—capturing spatial and element topologies. It enables a wide range of applications in early building design (Fig. \ref{Applications}) by extending Graph Neural Networks (GNNs) and KG-based large language models (KG-LLMs) \cite{Pan2024Unifying} to more flexible input formats. For instance, GNN-based performance prediction represents a state-of-the-art data-driven BPS approach, outperforming CNNs and FCNs \cite{Wu2025}, but its practical use has been limited by challenges in automatically interpreting building space networks; this framework overcomes that barrier. Similarly, LLM applications in building design have been constrained by building encoding methods. By supporting KG-based retrieval-augmented generation (KG-RAG), which improves LLM interpretability \cite{Li2026BuildingGPT}, the framework enables extension of IFC-based LLM applications to more flexible model formats, fostering broader and more creative applications in design intelligence.

To meet those scenarios, several convenience module is included in the framework for more diverse transformation, including the reversible transformation between *.obj/*.geo (a stream-based fast IO format designed for this framework), and the KG-based BIM model (*.rdf/*.owl) to LPG model.(*.xml/*.json) (1) *.obj transforming and cleansing to *.geo, a stream-based fast IO format designed for this framework; (2) *.geo or MoosasModel transforming TO/FROM *.rdf based KG, including/excluding 2LSB definition following IFC4.0; (3) Moosasmodel TO/FROM *.xml based space topology, and pruned LPG in *.json.

In addition, for applications based on Graph Neural Networks (GNN), we have incorporated a recent study proposing a Labelled Property Graph (LPG) representation that shows strong potential for building performance prediction tasks \cite{liGeometryGraphAutomation2026}. The proposed representation includes a data-alignment module that converts knowledge graphs (KG) into LPG structures using the BTO ontology, which constitutes part of the BIM ontology employed in our framework. By integrating this module into our transformation pipeline, the framework can be further extended to support GNN-based performance prediction, enabling graph-oriented machine learning analyses on the generated building knowledge graphs.

Regarding scalability, the proposed framework supports limited multi-building processing; however, its robustness decreases when multiple buildings with similar floor elevations are included in a single model (e.g., 2.9m, 3.2m, 3.5m). In such situations, the increased density of closely spaced building levels may introduce ambiguities in spatial boundary identification and topology extraction, thereby affecting the reliability of the ASG process. Consequently, the current framework is not well suited for ultra-large-scale models containing many buildings within a single geometry dataset.

For city-scale applications, a more practical strategy is to integrate hierarchical approach that decomposes the urban model into individual building units. For example, city models stored in *.shp format can be spatially partitioned and converted into separate building geometries (e.g., multiple *.obj files), which can then be processed independently within the DM→BEM pipeline. This building-level processing strategy can significantly improve robustness while enabling scalable analysis of large urban datasets.

\subsection{Case of AI Intervention}
To demonstrate the potential of AI-driven applications, this study incorporates a command-based question answering (QA) task by integrating knowledge graphs (KG) and large language models (LLMs) \cite{Pan2024Unifying}. The QA system is built upon the BIM generated from the design model of the real-world case study—the 210 Government Office Building in Beijing. Entities are extracted from the BIM to construct both a vector database and an entity graph, enabling a Vector-Graph Retrieval Augmented Generation (VG-RAG) framework \cite{Li2026BuildingGPT} to enhance QA performance.
	
\begin{figure}
	\centerline{\includegraphics[width=\columnwidth]{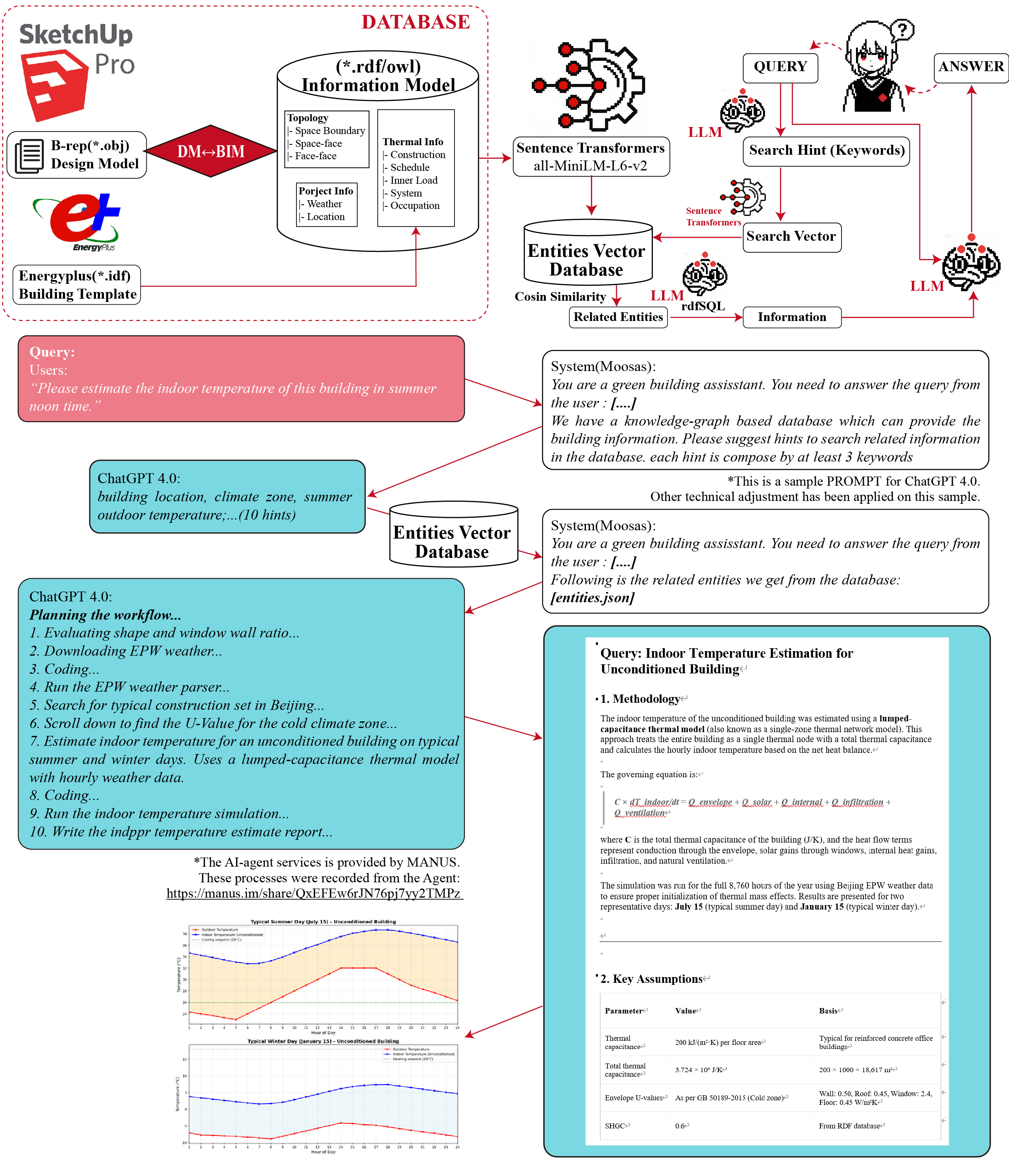}}
	\caption{Potential applications with this framework.}
	\label{AIApp}
\end{figure}

An end-to-end automatic pipeline was developed based on the MANUS AI-Agent service, including query analysis, information retrieval, planning, coding, and result generation. The agent first formulates an analysis plan based on KG retrieval, and then generates executable code to perform calculations or simulations. Figure~\ref{AIApp} illustrates the overall framework and an example query.

To evaluate performance and answer accuracy, three queries with increasing levels of complexity are designed:

\begin{itemize}
	
	\item \textbf{Query 1: “Please calculate the shape factor of this building.”}
		This query targets a deterministic value that can be manually verified (the reference shape factor is 0.275). By querying the KG, the agent successfully extracts key geometric attributes, including floor area, volume, façade area, and roof area. The generated response includes a complete calculation process—covering definitions, standard references, and intermediate steps—leading to the final result: \textit{“The exterior envelope area $F_0$ is 23,068.19~m$^2$, and the building volume $V_0$ is 83,776.73~m$^3$, resulting in a shape factor of 0.275~m$^{-1}$.”} This demonstrates high reliability in structured information extraction from the KG.
	
	\item \textbf{Query 2: “Please suggest construction assemblies for this building, including external/internal walls, glazing, floors, and roof.”}
		This query involves knowledge synthesis, requiring integration of KG-derived data and external regulatory information. The construction specifications are expected to comply with the Chinese standard \textit{GB55015-2021}. Additionally, an \texttt{*.idf} template embedded in the BIM must be interpreted and compared against regulatory requirements. The AI-generated construction assemblies are summarized in Table~\ref{tab:thermal_compliance}. While the agent demonstrates the ability to combine heterogeneous information sources, inconsistencies and insufficient detail are observed, primarily due to limitations in external information retrieval. This highlights the need for more reliable and domain-specific databases in construction-oriented AI applications.
	
	\item \textbf{Query 3: “Please estimate the indoor temperature of this building for typical summer and winter days under unconditioned conditions.”}
		This query requires multi-step reasoning and numerical estimation based on building properties and external weather data. The agent produces a structured report including methodology, assumptions, results, and a daily temperature profile (Fig.~\ref{AIApp}). The results indicate that the agent can effectively retrieve and integrate KG data with external weather information. However, the computational approach is limited to simplified methods (e.g., degree-day estimation), and cannot yet replace detailed building energy simulation. This suggests that integrating external simulation engines as callable functions would significantly enhance the analytical capability of AI agents.
	
\end{itemize}

In summary, the KG generated by the proposed framework effectively supports case-based reasoning and quantitative analysis, particularly for geometry- and topology-related queries. However, further improvements are required in external knowledge retrieval and simulation capabilities. Future work should focus on integrating reliable domain databases and simulation modules via function-calling mechanisms to enable more advanced and accurate AI-assisted building performance analysis.

\begin{table}[htbp]
	\centering
	\caption{Building Element Thermal Performance Compliance suggested by AI (W/m²K)}
	\label{tab:thermal_compliance}
	{ 
		\begin{tabular}{l l l l l}
			\toprule
			Building Element & \multicolumn{1}{l}{Suggested Construction} & \multicolumn{1}{l}{Calculated U-value } & \multicolumn{1}{l}{Code Requirement} & Compliant? \\  
			\midrule
			External Wall & \shortstack[l]{100mm XPS insulation \\ on RC wall}  & 0.26 & 0.50 & Yes \\
			\midrule
			Roof & \shortstack[l]{120mm XPS insulation \\ (inverted roof)} & 0.24 & 0.45 & Yes \\
			\midrule
			Glazing & \shortstack[l]{Double-glazed, Low-E, \\ Argon-filled} & $\sim$2.0 & 2.4 & Yes \\
			\midrule
			Ground Floor & \shortstack[l]{80mm XPS insulation \\ under slab} & 0.34 & 0.45 & Yes \\
			\bottomrule
		\end{tabular}
	}
	\\[1ex] 
	\raggedright\footnotesize *This table is generated by AI.
	\label{tab:building_thermal_compliance}
\end{table}

\section{Conclusion}
This paper presents a novel and robust geometry-transformation framework that bridges unstructured B-rep Design Models (DM), knowledge-graph-based Building Information Models (BIM), and EnergyPlus Building Energy Models (BEMs). The framework integrates six cleansing methods and three 1LSB Auto Space Generation (ASG) approaches, and systematically evaluates their combined performance.

Validation across three DM datasets demonstrates that the framework achieves a 100\% processing success rate, maintains high robustness with all cleansing modules enabled, and produces BIM KGs with an average space-related topology accuracy of 0.871 (nGED). The subsequent BIM-to-BEM transformation achieves high fidelity in hourly energy simulations, with $R^2 = 0.946$ for heating load and $R^2 = 0.987$ for cooling load.

By enabling fully automatic transformations from DM to BIM and BEM, the framework extends LPG- and KG-based BIM techniques and AI intervention to unstructured design models, unlocking a range of novel applications. Many imaginative applications can thus be realized: such as the KG-LLM based model interpretation DIRECTLY on a B-rep model, the BIM-based Building Performance Optimization (BPO) on ANY sketch design, and quick and simple BPS on those DMs. Besides, a new UBEM workflow can be discussed with the sequence of SHP-OBJ-BIM-BEM, which can process more complex urban contents with elaborated space structure.

Despite these advancements, future work will focus on improving the computational efficiency of cleansing routines, mitigating geometric errors from constructive operations, and implementing advanced thermal zoning algorithms. These developments aim to strengthen the framework’s utility in early-stage building design.

\section{Data Availability}
	The Moosas package and testing code is available through:
	https://github.com/KLEGB/moosas
	A reproducible example is available via Code Ocean capsule:
	https://doi.org/10.24433/CO.2302305.v2
\section{Acknowledgments}
	The study was supported by the National Natural Science Foundation of China (Grant No.52425801, 52394223 and 52130803).
\bibliographystyle{elsarticle-num} 

\bibliography{ref}



\end{document}